\documentclass[iop,twocolappendix]{emulateapj}
\usepackage{natbib}
\usepackage{psfig}
\usepackage{CJK}

\newcommand{\pdif}[2]{\ensuremath{ \frac{\partial #1}{\partial #2}}}

\begin{document}

\begin{CJK*}{UTF8}{gbsn}

\title{Radiation Feedback in ULIRGS:  Are Photons Movers and Shakers?}
\author{Shane W. Davis\altaffilmark{1}, Yan-Fei Jiang(姜燕飞)\altaffilmark{2,3}, James M. Stone\altaffilmark{4}, and Norman Murray\altaffilmark{1}}
\altaffiltext{1}{Canadian Institute for Theoretical Astrophysics. Toronto, ON M5S3H4, Canada}
\altaffiltext{2}{Harvard-Smithsonian Center for Astrophysics, 60 Garden Street, Cambridge, MA 02138, USA}
\altaffiltext{3}{Einstein Fellow}
\altaffiltext{4}{Department of Astrophysical Sciences, Princeton University, Princeton, NJ 08544, USA}

\begin{abstract}

We use our variable Eddington tensor (VET) radiation hydrodynamics
code to perform two-dimensional simulations to study the impact of
radiation forces on atmospheres composed of dust and gas.  Our setup
closely follows that of Krumholz \& Thompson, assuming that dust and
gas are well-coupled and that the radiation field is characterized by
blackbodies with temperatures $\gtrsim 80$ K, as might be found in
ultraluminous infrared galaxies.  In agreement with previous work, we
find that Rayleigh-Taylor instabilities develop in radiation supported
atmospheres, leading to inhomogeneities that limit momentum exchange
between radiation and dusty gas, and eventually providing a near
balance of the radiation and gravitational forces.  However, the
evolution of the velocity and spatial distributions of the gas differs
significantly from previous work, which utilized a less accurate
flux-limited diffusion (FLD) method.  Our VET simulations show
continuous net acceleration of the gas, with no steady-state reached
by the end of the simulation.  In contrast, FLD results show little
net acceleration of the gas and settle in to a quasi-steady, turbulent
state with low velocity dispersion.  The discrepancies result
primarily from the inability of FLD to properly model the variation of
the radiation field around structures that are less than a few optical
depths across.  We conclude that radiation feedback remains a viable
mechanism for driving high-Mach number turbulence.  We discuss
implications for observed systems and global numerical simulations of
feedback, but more realistic setups are needed to make robust
observational predictions and assess the prospect of launching
outflows with radiation.

\end{abstract}


\section{Introduction}

Observations of star-formation in the Milky way and other galaxies
provide consistent evidence that some feedback mechanism or mechanisms
hamper the collapse of interstellar gas to form stars.  For example,
The Kenicutt-Schmidt law \citep{Kennicutt1998} implies that, on
average, only a few percent of the available gas actually collapses to
form stars per dynamical time.  Observations of molecular gas in
rapidly star-forming ultraluminous infrared galaxies (ULIRGs) indicate
that turbulent velocities of up to $\sim 100 \; \rm km \; s^{-1}$ are
present \citep[e.g.][]{DownesSolomon1998}.  And most dramatically,
galaxy scale outflows of cold, neutral gas are inferred in galaxies
ranging from nearby dwarf starbursts to ULIRGs and rapidly
star-forming galaxies at high redshift
\citep[e.g.][]{Heckmanetal1990,Pettinietal2001,SchwarzMartin2004,Rupkeetal2005}.

Although a number of promising mechanisms have been proposed to
explain these observations, we restrict our attention to the potential
role of radiation pressure on dust grains in driving turbulence,
hampering gravitational collapse, and launching outflows in such
environments
\citep{Scoville2001,Murrayetal2005,Thompsonetal2005}. This possibility
has been explored extensively in recent years, with a number of
studies considering how (in)effective radiation driving may be in
various environments
\citep[e.g.][]{KrumholzMatzner2009,AndrewsThompson2011,Hopkinsetal2011,Hopkinsetal2012,Wiseetal2012,KrumholzThompson2012,KrumholzThompson2013,SocratesSironi2013}

Here we focus specifically on the implications of the Rayleigh-Taylor
instability \citep[hereafter RTI; see e.g.][]{Chandrasekhar1961}.  In
recent work, \citet[ hereafter KT12]{KrumholzThompson2012} and
\citet{KrumholzThompson2013} have argued that the RTI may play a
significant role in limiting the exchange of momentum between
radiation and dusty gas, ultimately reducing the role of radiation
feedback in star-forming environments. A general, analytical
calculation of the linear growth of the radiative RTI does not exist
\citep[see
  e.g.][]{MathewsBlumenthal1977,Krolik1977,JacquetKrumholz2011,Jiangetal2013},
and non-linear evolution can generally only be explored via numerical
simulations \citep[KT12;][]{Jiangetal2013}.

In this paper, we attempt to replicate the results of KT12, who
numerically solve the equations of radiation hydrodynamics using an
implementation of the flux-limited diffusion algorithm in the ORION
code \citep{Krumholzetal2007}.  We utilize both a variable Eddington
tensor method \citep{Davisetal2012,Jiangetal2012} and our own version
of the flux-limited diffusion method (Jiang et al., in preparation),
each of which are implemented as part of the Athena astrophysical
fluid dynamics code \citep{Stoneetal2008}.  Although we reproduce some
aspects of the KT12 results, we find significant discrepancies that
arise from innacuracies of the flux-limited diffusion treatment.

The plan of this work is as follows:  We review the equations solved
and summarize our numerical methods in section \ref{equations}.  We
describe the setup of our numerical simulations, including
formulations for the opacities, boundary conditions, and initial
conditions in section \ref{setup}.  We summarize our key results from
our numerical simulation in section \ref{results} and discuss their
implications in section \ref{discussion}.  We provide our
conclusions in section \ref{summary}.

\section{Equations and Numerical Methods}
\label{equations}

In this work we solve the equations of radiation hydrodynamics using
the variable Eddington tensor (VET) implementation
\citep{SekoraStone2010,Davisetal2012,Jiangetal2012}.  The primary set
of equations to be solved are the coupled systems of hydrodynamics and
the radiation moment equations.  The standard fluid equations are
solved using a finite volume formulation as discussed in
\cite{Stoneetal2008}.  We solve the radiation system in the mixed
frame approach \citep[see e.g.][]{MihalasMihalas1984}, where the
radiation moments are Eulerian frame quantities, while the
emissivities and opacities correspond to the comoving frame quantities
\citep[c.f.][]{MihalasKlein1982}.  We follow the derivation of
\cite{Lowrieetal1999}, which retains all terms of order $v/c$ and some
terms of order $(v/c)^2$.  The equations correspond to mass
conservation
\begin{equation}
\pdif{\rho}{t} + \mathbf{\nabla} \cdot \left(\rho \mathbf{v} \right) = 0 \label{eq:mass} ,
\end{equation}
momentum conservation
\begin{equation}
\pdif{\left(\rho\mathbf{v}\right)}{t} + \mathbf{\nabla} \cdot \left( \rho\mathbf{v} 
\mathbf{v} + {\sf P}\right) = \rho \mathbf{g} - \mathbf{S}_r(\mathbf{P}),\label{eq:momentum}
\end{equation}
and energy conservation
\begin{equation}
\pdif{E}{t} + \mathbf{\nabla} \cdot \left(E \mathbf{v} + {\sf P} \cdot \mathbf{v}\right) = 
\rho \mathbf{g} \cdot \mathbf{v} -c S_r(E).
\label{eq:energy}
\end{equation}
In the above, $\rho$ is the gas density, $\mathbf{v}$ is the fluid
velocity, $\mathbf{g}$ is the gravitational acceleration, and $E$ is
the total (fluid) energy density $E=p/(\gamma-1)+\rho v^2/2$.  The
pressure tensor $\sf P$ is defined as ${\sf P} = p {\sf I}$, where $p$
is the gas pressure and $\sf I$ is the identity matrix.

The quantities $\mathbf{S}_r(\mathbf{P})$ and $S_r(E)$ are the radiation
momentum and energy source terms, respectively.  They are given by
\begin{eqnarray}
\mathbf{S}_r(\mathbf{P}) & = &-\sigma_{F}\left[\mathbf{F}_r-
\left(\mathbf{v} E_r+\mathbf{v} \cdot {\sf P}_r\right)\right]/c  \nonumber \\ 
& &+\mathbf{v}(\sigma_{\rm P}a_rT^4-\sigma_{E}E_r)/c,\label{eq:srcrade}
\end{eqnarray}
and
\begin{eqnarray}
S_r(E) & = & (\sigma_{\rm P}a_rT^4-\sigma_{E}E_r)  \nonumber \\
 & & +\sigma_{F}\frac{\mathbf{v}}{c^2}\cdot\left[\mathbf{F}_r-
\left(\mathbf{v} E_r+\mathbf{v}\cdot{\sf P} _r\right)\right].\label{eq:srcradp}
\end{eqnarray}
Here $E_r$ is the radiation energy density, $\mathbf{F}_R$ the
radiation flux, ${\sf P}_r$ is the radiation pressure tensor, $a_r$
is the radiation constant, $T$ is the gas temperature, $\sigma_{E}$ is
energy mean opacity, $\sigma_{\rm P}$ is the Planck mean opacity, and
$\sigma_{F}$ is the flux mean opacity.  In this work we neglect any
scattering contribution to the opacity. The radiation source terms on the right hand side of
equations (\ref{eq:momentum}) and (\ref{eq:energy}) are included via a modified
Godunov method, as described in \citet{Jiangetal2012}.

In order to compute these quantities, we solve the radiation moment equations
representing conservation of radiation energy
\begin{eqnarray}
\pdif{E_r}{t} + \mathbf{\nabla} \cdot \mathbf{F}_r=cS_r(E),\label{eq:radenergy}
\end{eqnarray}
and conservation of radiation momentum
\begin{eqnarray}
\frac{1}{c^2}\pdif{\mathbf{F}_r}{t}+\mathbf{\nabla} \cdot{\sf P}_r=\mathbf{S}_r(\mathbf{P})\label{eq:radmomentum}.
\end{eqnarray}
This radiation system is solved using a backward-Euler formulation
\citep{Jiangetal2012}.

\subsection{The Short-Characteristics Method}
\label{shortchar}

The above equations are incomplete because there is no evolution equation
for the radiation pressure tensor.  We address this by computing
the eponymous variable Eddington tensor ${\sf f} \equiv {\sf P}_r/E_r$.
We approximate ${\sf f}$ by solving the time-independent equation of
radiation transfer of the form
\begin{equation}
\hat{n} \cdot \nabla I = \sigma_{F} \left(\frac{a_r c}{4\pi} T^4 - I\right),
\label{eq:radtrans}
\end{equation}
where $I$ is the specific intensity for an angle defined by the unit
vector $\hat{n}$.  Hence, we have adopted a grey opacity treatment in
which the characteristic opacity is assumed to correspond to the flux
mean opacity.  For the computation of the $\sf f$, we make no
distinction between the comoving and Eulerian frames, since equation
(\ref{eq:radtrans}) drops all order $v/c$ terms or higher, including a
$c^{-1}\partial I/\partial t$ term that would otherwise be present.
The neglect of the $c^{-1}\partial I/\partial t$ term is typically a
good approximation whenever $v/c \ll 1$ but the neglect of
velocity-dependent terms generally requires $\tau v/c \ll 1$ if $\tau$
is a characteristic optical depth for the system.  In the results
presented here $v_{\rm max}/c \sim 10^{-4}$ and $\tau_{\rm max} \sim
15$, so the neglect of these terms is a good approximation.

On each timestep, this equation is solved using the method of short
characteristics \citep{KunaszAuer1988}, as described in
\citet{Davisetal2012}.  As the radiation transfer calculation
proceeds, the mean intensity $J$, the first moment $\mathbf{H}$, and
the second moment $\sf K$ are tabulated in each grid zone.  These are
then used to compute ${\sf f} ={\sf K}/J$ for the subsequent timestep.
We integrate the radiation field along 84 rays, distributed nearly
uniformly over the half unit sphere, taking advantage of the
reflection symmetry of two dimensional Cartesian domains.  Further
discussion of the angular grid and impact of angular resolution is
provided in the appendix.

\subsection{The Flux-limited Diffusion Method}

We have also implemented a flux-limited diffusion (FLD) algorithm in
Athena. Here, we only briefly describe the equations and
implementation.  A more detailed description of the algorithm and its
implementation will be provided in a forthcoming work (Jiang et al.,
in preparation).

In FLD, one drops the radiation momentum equation from the evolution
equations, replacing it with a diffusion equation for the flux
\begin{equation}
\mathbf{F}_r=-\frac{c \lambda}{\sigma_{F}} \nabla E_r,\label{eq:fld}
\end{equation}
where $\lambda$ is the flux-limiter.  The radiation energy equation is
then solved by substituting this relation for $\mathbf{F}_r$ in
equations (\ref{eq:srcrade})-(\ref{eq:radenergy}).  Following
\citet{LevermorePomraning1981}, we assume a flux limiter of the form
\begin{eqnarray}
\lambda &= &\frac{1}{R}\left(\coth{R} - \frac{1}{R}\right)\nonumber \\
R &=&\frac{| \nabla E_r |+\beta}{\sigma_{F} E_r}.\label{eq:limiter}
\end{eqnarray}
To specify ${\sf P}_r$ we use the Eddington tensor
\begin{eqnarray}
{\sf f} =  \frac{1}{2}(1-f){\sf I}+\frac{1}{2}(3 f-1)\hat{f}\hat{f} 
\end{eqnarray}
with $f = \lambda+ \lambda^2 R^2$ and $\hat{f} = \nabla E_r/|\nabla E_r|$.

Following \citet{ShestakovOffner2008} we have added a $\beta$
parameter to the definition of $R$, which acts as an effective floor
on $R$ in optically thin regions. This addition was necessary to
robustly obtain convergence in our backward-Euler scheme, in which one
needs to solve a large matrix equation.  We find poor
convergence for our multigrid solver unless $\beta \gtrsim 10^{-4}$.
These equations are implemented in a manner analogous to the VET
method described above.  The source terms are coupled to the hydro
equations using our modified Godunov method and the radiation energy
equation is solved using a backward-Euler algorithm.  Finally, we note
that $\mathbf{F}_r$ is the Eulerian frame flux in the above equations.
Strictly speaking, it is the comoving frame flux for which the
diffusion approximation applies in high optical depth media.  However,
the low vales of $v/c$ in the simulations lead to very small
differences between the Eulerian and comoving frame fluxes.  

Due to the ad hoc aspects of this approximation, it is not as reliable
as our VET method, which evolves the radiation momentum equation and
uses a solution of the radiation transfer equation to estimate $\sf
f$.  Modulo the velocity dependent terms that we neglect in computing
$\sf f$ (and only in computing $\sf f$), the accuracy of our
approximation can always be improved by increasing the angular
resolution.  In contrast, FLD will {\it always be inaccurate} at some
level in regions of the flow where the characteristic optical depths
are near unity or less.\footnote{Although FLD is often claimed to be
  accurate in the optically thin limit, it assumes $|\mathbf{F}_r|
  \rightarrow E_r c$, which only applies to a radiation field with
  perfectly parallel rays. This is typically only obtained at large
  distances from a point source emitter and is not general.}
Well known deficiencies of this method include its inability to cast
shadows \citep[see e.g.][]{HayesNorman2003} because the radiation
field diffuses around opaque barriers, even in optically thin
environments.  We have implemented FLD in Athena only to facilitate
comparison of our VET calculation with previous FLD-based results,
since it aids in determining which aspects of the calculation depend
on the transfer method.

\section{Simulation Setup}

\label{setup}

Our simulation setup follows the computations performed in KT12.  The
goal is to simulate the evolution of the interstellar gas in ULIRGs,
although the results may generalize to other rapidly star forming
environments.  We assume that the gas and dust are strongly coupled,
both dynamically and thermally.  This assumption should be quite
reasonable for ULIRGs, as discussed in appendix A of
\citet{KrumholzThompson2013}.  Hence, the gas and dust are assumed to
share a common temperature (hereafter simply referenced as the gas
temperature $T$), and any momentum exchanged between the dust and
radiation field is rapidly shared via collisions with the gas.

Following KT12, the Planck $\kappa_{\rm P}$ and Rosseland $\kappa_{\rm
  R}$ mean opacities are given by
\begin{eqnarray}
\kappa_{\rm P} & = & 0.1 \left( \frac{T}{10 \; K}\right)^2 \; {\rm cm^2 \; g^{-1}}\nonumber\\
\kappa_{\rm R} & = & 0.0316 \left( \frac{T}{10 \; K}\right)^2 \; {\rm cm^2 \; g^{-1}}.\label{eq:opacity}
\end{eqnarray}
This $\kappa \propto T^2$ scaling approximately holds for dust in
thermal equilibrium with a blackbody radiation field for $T \lesssim
150 K$ \citep{Semenovetal2003}.  This implicitly assumes the
characteristic temperature of the radiation and $T$ are equal.  Since
we evolve the radiation energy density, we can define a characteristic
radiation temperature $T_r=(E_r/a_r)^{1/4}$ and check a posteriori if
$T_r \simeq T$.  This is generally a good approximation although
modest differences are present in some regions, which motivated us to
perform alternative simulations with $\kappa_{\rm R,P} \propto T_r^2$.
The two sets of simulations agreed very closely with each other so
we report only the $\kappa_{\rm R,P} \propto T^2$ here.  These
assumptions imply that the radiation field is characterized by
blackbodies with $T_r \sim T \sim 80 - 200$ K.  Since the main source
of radiation is massive stars emitting primarily in the UV, this
analysis assumes the direct stellar radiation has already been
reprocessed by the dust and converted to infrared radiation.

The simulations are performed on a two-dimensional, Cartesian grid.
For simplicity, the gravitational acceleration $g$ is assumed to be
constant and the radiation field is sourced by a constant flux $F_*$,
incident at the lower boundary.  This flux defines a characteristic
temperature $T_*=(F_*/a_r c)^{1/4}$, from which we can define a
characteristic sound speed $c_*^2=k T_*/(\mu m_H)$, scale height
$h_*=c_*^2/g$ and sound crossing time $t_*=h_*/c_*$. Following KT12,
we assume $\mu=2.33$ and choose $T_*=82$ K, corresponding to $F_*
=2.54 \times 10^{13} \; \rm L_\odot \; kpc^{-2}$ and $\kappa_{\rm
  R,*}=2.13 \; \rm cm^2 \; g^{-1}$.  The remaining free parameters are
$g$ and the surface mass density $\Sigma$, which can be specified in
terms of the dimensionless Eddington ratio\footnote{Note that this
constitutes only a local assessment of the force balance.  Since $g$
varies with height and radius in real systems, a local balance between
radiation and gravitational forces simply means the disk is radiation
pressure supported.  The global Eddington ratio of the system may be
much less than unity.}
\begin{equation}
f_{\rm E,*} = \frac{\kappa_{\rm R,*} F_*}{g c}
\end{equation}
and optical depth
\begin{equation}
\tau_* = \kappa_{\rm R,*} \Sigma.
\end{equation}

\subsection{Initial Conditions}
\label{init}

As in KT12, we assume an initially static and isothermal atmosphere
with $T=T_*$.  The density is initialized as an exponential profile
with scale height $h_*$, which would correspond to hydrostatic
equilibrium if radiation pressure support was negligible.  For both
the initial condition and subsequent evolution we impose a density
floor of $10^{-10} \rho_*$, where $\rho_* = \Sigma/h_*$ is the maximum
initial density at the lower boundary.  We assume $T_r=T$ everywhere,
which gives a constant $E_r=a_r T_*^4=c F_*$.  Hence, for even
moderate values of $\tau_*$ and $f_{\rm E,*}$, this initial condition
is neither in thermodynamic nor hydrostatic equilibrium.

On top of this initial density profile we include a perturbation of
the general form
\begin{equation}
\frac{\delta \rho}{\rho} = 0.25 (1 \pm \chi) \sin{(2 \pi x/\lambda)},\label{eq:pert}
\end{equation}
where $\lambda = 0.5 L_x$, and $L_x$ is the domain width. Here, $\chi$
is identically zero for sinusoidal perturbations or can be a random
number uniformly distributed between -0.25 and 0.25. If $\chi=0$, the
perturbation is identical to KT12, but we found it useful to
consider simulations with random perturbations on top of the
sinusoidal variation.

Simulations parameters for our primary simulations are summarized in
Table \ref{t:parameters}.

\begin{deluxetable*}{lccccccc}
\tablecolumns{8}
\tablecaption{Simulation Summary\label{t:parameters}}
\tablehead{
\colhead{Label} & 
\colhead{Method} & 
\colhead{Perturbation} & 
\colhead{$\tau_*$} & 
\colhead{$f_{{\rm E},*}$} & 
\colhead{$(L_x \times L_y)/h_*$} & 
\colhead{$N_x \times N_y$} &
\colhead{$t_{\rm end}/t_*$}
}
\startdata
T10F0.02VET & VET & sin & 10 & 0.02 & $256 \times 128$ & $512 \times 256$ & 80\\
T3F0.5VET & VET & sin/random & 3 & 0.5 & $256 \times 512$\tablenotemark{a} & $512 \times 1024$\tablenotemark{a} & 158\\
T3F0.5FLD & FLD & sin/random & 3 & 0.5 & $256 \times 512$ & $512 \times 1024$ & 200\\
\enddata
\tablenotetext{a}{Simulation T3F0.5VET was restarted at $80 t_*$ in a domain with $L_y=1024$ and $N_y=2048$.}
\end{deluxetable*}

\subsection{Boundary Conditions}

For all simulations, both the radiation and hydrodynamic quantities
obey periodic boundary conditions in the horizontal ($x$-coordinate)
direction.  For the hydrodynamic quantities, we impose reflecting
boundary conditions at the lower vertical ($y$-coordinate) boundary
and outflow boundary conditions on the upper vertical boundary.  These
boundary conditions match those of KT12, modulo implementation
differences between Athena and ORION.  The exception is that we simply
continue $\rho$ and $E$ into the ghost zones at our upper boundary
whereas KT12 fix $\rho=10^{-13} \rho_*$ and $T=10^3 T_*$ there.

For the lower boundary condition of the VET runs, we apply reflecting
boundary conditions on $F_{rx}$ and require
\begin{equation}
F_{ry}=F_*+v_y E_r+(\mathbf{v} \cdot {\sf P}_r)_y.
\end{equation}
Hence, the comoving flux in the ghost zones is equal to $F_*$.  We
specify $E_r$ by integrating the time-independent vertical momentum
equation, assuming the Eddington tensor at the lower boundary of the
computational domain applies throughout the ghost zones.  At the upper
boundary, both components of $\mathbf{F}_{r}$ are continued in to the
ghost zones and we specify $E_r$ using
\begin{equation}
E_r=\frac{J}{H_y} F_{ry},
\end{equation}
where $\mathbf{H}$ and $J$ are the first moment and mean intensity
(respectively) returned by the short-characteristics module. In other
words, the ratios of the vertical component of the first moment to the
``zeroth'' moment for the two calculations are forced to match
identically in the ghost zones.

For the lower boundary condition of the FLD computations, we compute
the flux limiter using equation (\ref{eq:limiter}) and solve for $E_r$
using equation (\ref{eq:fld}), assuming $F_{ry}=F_*$ and no horizontal
gradient in energy density.  At the upper boundary we assume a fixed
$E_r=a T_*^4$ in the ghost zones for consistency with KT12.

For the lower boundary in the short-characteristics module, we
integrate the outgoing (downward) radiation intensity over angle
$J_-$.  We then assume isotropic incoming (upward) radiation. The
intensity of the incoming radiation is normalized so that the
corresponding angle integrated upward intensity $J_+$ obeys
$H_y=J_+-J_-=F_*$.  At the upper boundary we assume that the intensity
of incoming radiation is zero.  Periodic boundary conditions in the
horizontal direction are imposed by iterating the
short-characteristics solution to convergence, as described in
\citet{Davisetal2012}.

\section{Results}
\label{results}

KT12 computed the equilibrium profiles for one-dimensional atmospheres
obeying the assumptions of section \ref{setup}, assuming equation
(\ref{eq:fld}) holds.  They find that for a given value of $\tau_*$,
there is a maximum value for $f_{\rm E,*}$ ($f_{\rm E,crit}$) for
which their iterative method converges.  Above this value, no
equilibrium solution exists because the radiation and gravitational
forces cannot be forced to match exactly due to the temperature
dependence of the opacities. Their results for $\kappa \propto T^2$
are summarized in their figure 2.  Although their results are derived
using the FLD approximation, we expect that their boundary curve
should approximately demarcate the locus of stable equilibria
solutions for our VET calculations, as well.

Here we consider two sets of $\tau_*$ and $f_{\rm E,*}$, as summarized
in Table \ref{t:parameters}.  The first ($\tau_*=10$, $f_{\rm E,*}=0.02$)
falls in the range where an equilibrium exists, while the second
($\tau_*=3$, $f_{\rm E,*}=0.5$) falls in the unstable regime.  We
choose these parameters to match simulations performed by KT12.

\subsection{Stable Case}
\label{stable}

We first consider the T10F0.02VET run, which we run using our VET
module.  The parameters for this run place it in the stable
regime, according to the one-dimensional equilibrium solutions and
two-dimensional simulation results of KT12.  We initialized the
simulation with a sinusoidal perturbation only (i.e no random
fluctuations), as described in section \ref{init} and ran it for $80
t_*$.

\begin{figure}
\includegraphics[width=0.5\textwidth]{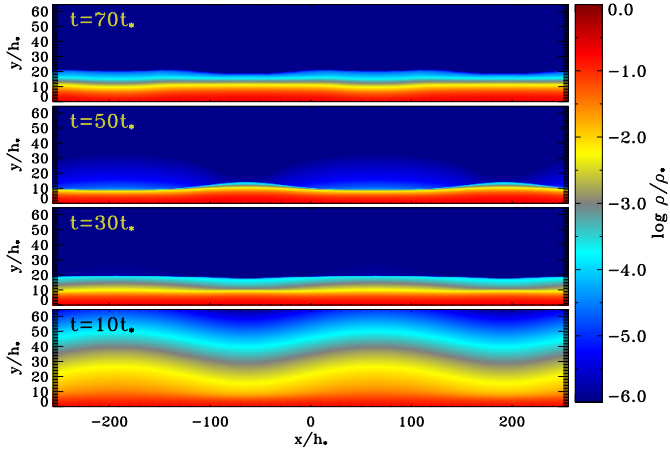}
\caption{Density distribution for four snapshots from the T10F0.02VET
  simulation.  Each panel shows the bottom quarter ($z/h_* \le 64$) of
  the domain.  The times correspond to $t/t_*=10$, 30, 50, and 70, as
  labeled.  The atmosphere initially expands upward, but falls back
  and then oscillates, similar to the KT12 results for the same
  parameters.
\label{f:dt10f0.02}}
\end{figure}

Figure \ref{f:dt10f0.02} shows four snapshots of the density from this
run.  At the beginning of the run, the large optical depth and
constant incident flux require $E_r$ to increase.  Since the radiation
and dusty gas are well-coupled, the increase in $T_{\rm rad}$ drives a
corresponding rise in $T$.  This leads to an increase in the opacity
and a corresponding rise in the optical depth and radiation force.  As
a result, the atmosphere briefly expands upward, relaxes, and begins
to oscillate around a quasi-equilibrium state.

\begin{figure}
\includegraphics[width=0.5\textwidth]{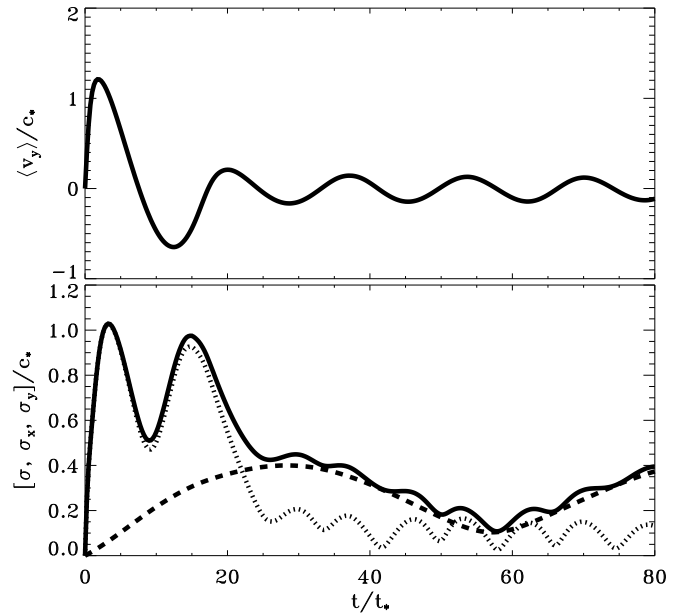}
\caption{Top panel: Mass-weighted mean velocity versus time for the
  T10F0.02VET simulation.  Bottom panel: Mass-weighted velocity
  dispersion versus time for the same run.  The curves correspond to
  $\sigma$ (solid), $\sigma_x$ (dashed), and $\sigma_y$ (dotted). In
  both panels, the velocities are normalized to the initial isothermal
  velocity $c_* =0.54 \; \rm km \; s^{-1}$.  After a transient
  acceleration, the atmosphere settles into slowly damped oscillatory
  behavior.
\label{f:sigt10f0.02}}
\end{figure}

This transient evolution and subsequent oscillatory behavior are most
clearly seen in the mass-weighted mean velocity
\begin{equation}
\langle \mathbf{v} \rangle = \frac{1}{M}\int \rho \mathbf{v} dV,
\end{equation}
and mass-weighted velocity dispersion
\begin{equation}
\sigma^2_{i} = \frac{1}{M} \int \rho (v_i-\langle v_i\rangle)^2 dV,
\end{equation}
where $M=\int \rho dV$ is the total mass in the atmosphere.  The top
panel of figure \ref{f:sigt10f0.02} shows $\sigma_x$ and $\sigma_y$,
along with the total velocity dispersion
($\sigma=(\sigma_x^2+\sigma_y^2)^{1/2}$) and the bottom panel shows
$\langle v_y \rangle$.

The initial transient acceleration lead to growth in both the vertical
and horizontal velocity dispersion, although the vertical component
initially grows faster.  After $\sim 3 t_*$, both $\langle v_y\rangle$
and $\sigma_y$ have already reached their maximum for the simulation.
After another $\sim 10 t_*$, both quantities settle down into
oscillations, with $\langle v_{y} \rangle$ varying about zero.  The
horizontal velocity dispersion is also oscillatory, albeit on a longer
timescale with a period greater than $50 t_*$.  Comparison with the
top panel of figure 7 in KT12 shows close agreement between the
results of the two different codes.

At the end of the run, the simulation is still oscillating, albeit
with a slow decay in $\langle v_y\rangle$. Reductions in the kinetic energy due
to ``numerical viscosity'' and diffusive damping of compressive
motions by the radiation field will eventually damp the oscillatory
behavior and the simulation will presumably settle into a steady
state.  However, this would seem to require a significantly longer
run time and we are not interested in the detailed properties of
equilibrium.  We stop the run at $80 t_*$, which is long enough to
confirm that we reproduce the KT12 results with our VET formalism for
this stable regime.

\subsection{Unstable Case}

We now focus on the unstable regime, considering simulations with
$\tau_*=3$ and $f_{\rm E,*}=0.5$.  This run has the lowest
ratio of $f_{\rm E,*}/f_{\rm E,crit}$ considered in KT12 for the
unstable regime.  Furthermore, it had the weakest transient
acceleration at the beginning of the simulation, reached the lowest
maximum vertical extent, and had the lowest maximum velocity
dispersion.  In this sense, the run represented the worst case for
driving turbulence and outflows among the runs considered by KT12.

\begin{figure*}[ht]
\includegraphics[width=\textwidth]{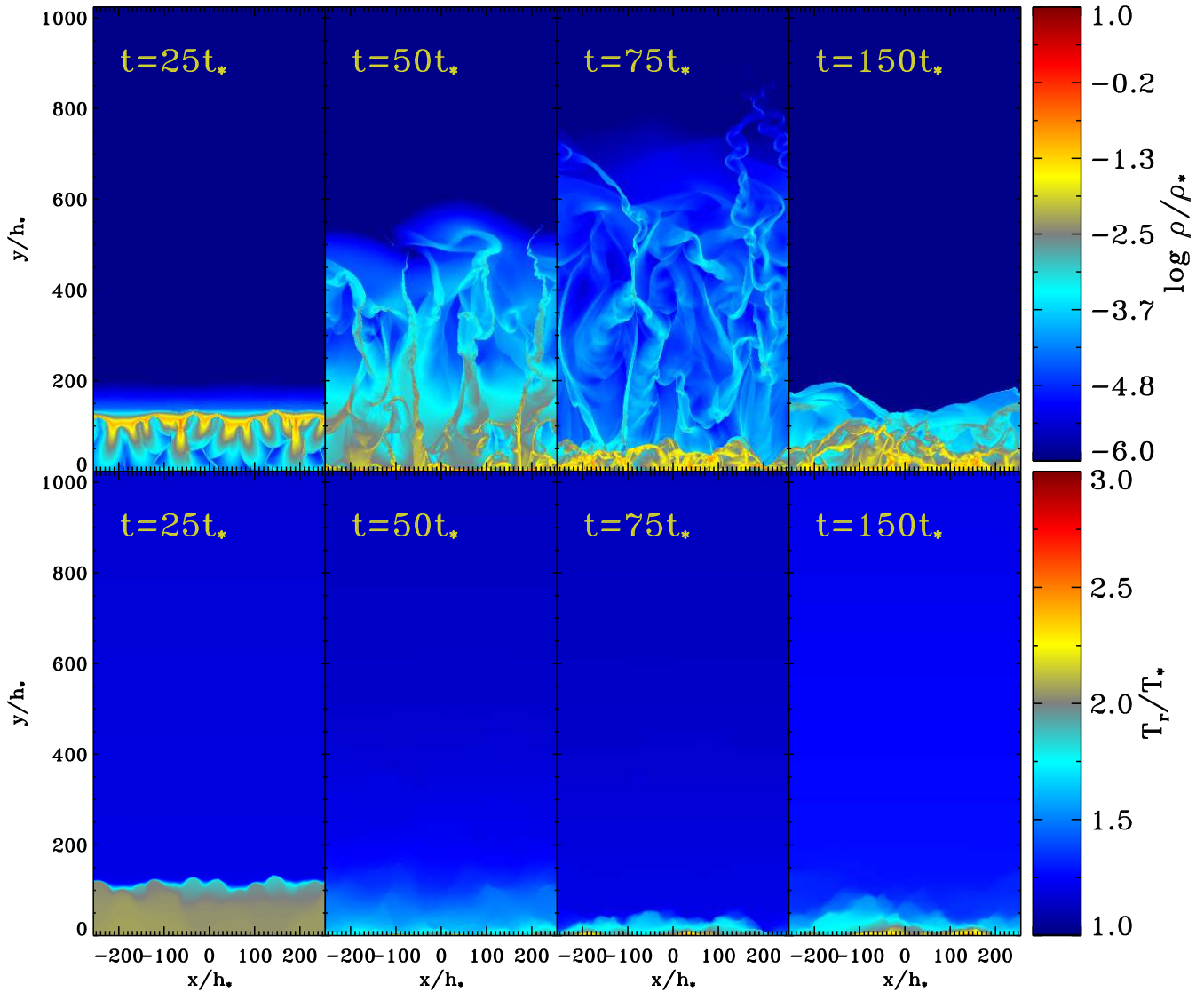}
\caption{Top panels: Density distribution for four snapshots
  from the T3F0.5FLD simulation.  Each panel shows the full simulation
  at the labeled times.  Bottom panels: Radiation temperature $T_r$
  distribution computed from the same snapshots as in the top row.
  The atmosphere heats up and is accelerated vertically as shell.  The
  shell becomes unstable, breaks up and mass falls back to the bottom
  of the domain, reaching a quasi-steady-state of turbulence at late
  times.
\label{f:t3f0.5fld}}
\end{figure*}

In our work, we consider two runs: one using our FLD module
(T3F0.5FLD) and one using the VET module (T3F0.5VET), in order to
assess the impact of the radiation transfer method.  In principle, we
could simply compare our VET method directly against the ORION FLD
results, but we perform Athena FLD runs to control for differences in
the hydrodynamics algorithms between Athena and ORION.  The
simulations were initialized with the goal of nearly reproducing the
KT12 setup.  One notable exception is that both runs have sinusoidal
and random perturbations, as described in equation (\ref{eq:pert}).
We made this choice after running a simulation with the VET module
that only included sinusoidal perturbations.  In that case, the
development of non-linear structure (due to the RTI, as discussed
below) was slower than seen in KT12 and most of the mass in the shell
was able to reach the top of the domain before this structure produced
a significant feedback on the radiative acceleration of the
shell. Since the sinusoidal perturbation is a artificial construction
and because we were interested in the evolution of an atmosphere with
significant non-linear structure, we seeded the RTI with additional
random perturbations.  This allowed for faster development of the RTI
in the VET runs.  We also initialize the FLD run with random
perturbations to facilitate direct comparison between our FLD and VET
runs.

We first consider the results from T3F0.5FLD, which we run for $200
t_*$.  This is run in a taller box than the T10F0.02VET simulation,
with the initial condition and boundary conditions described in
section~\ref{setup}.  Note that we began the simulations with
$\beta=10^{-4}$ in equation (\ref{eq:limiter}) to place a floor on $R$
and avoid convergence problems in our backward Euler scheme.  This was
sufficient for the first $\sim 100 t_*$, but we had to increase
$\beta=4 \times 10^{-4}$ at $t=100 t_*$ to ensure convergence
thereafter.

\begin{figure*}[ht]
\includegraphics[width=\textwidth]{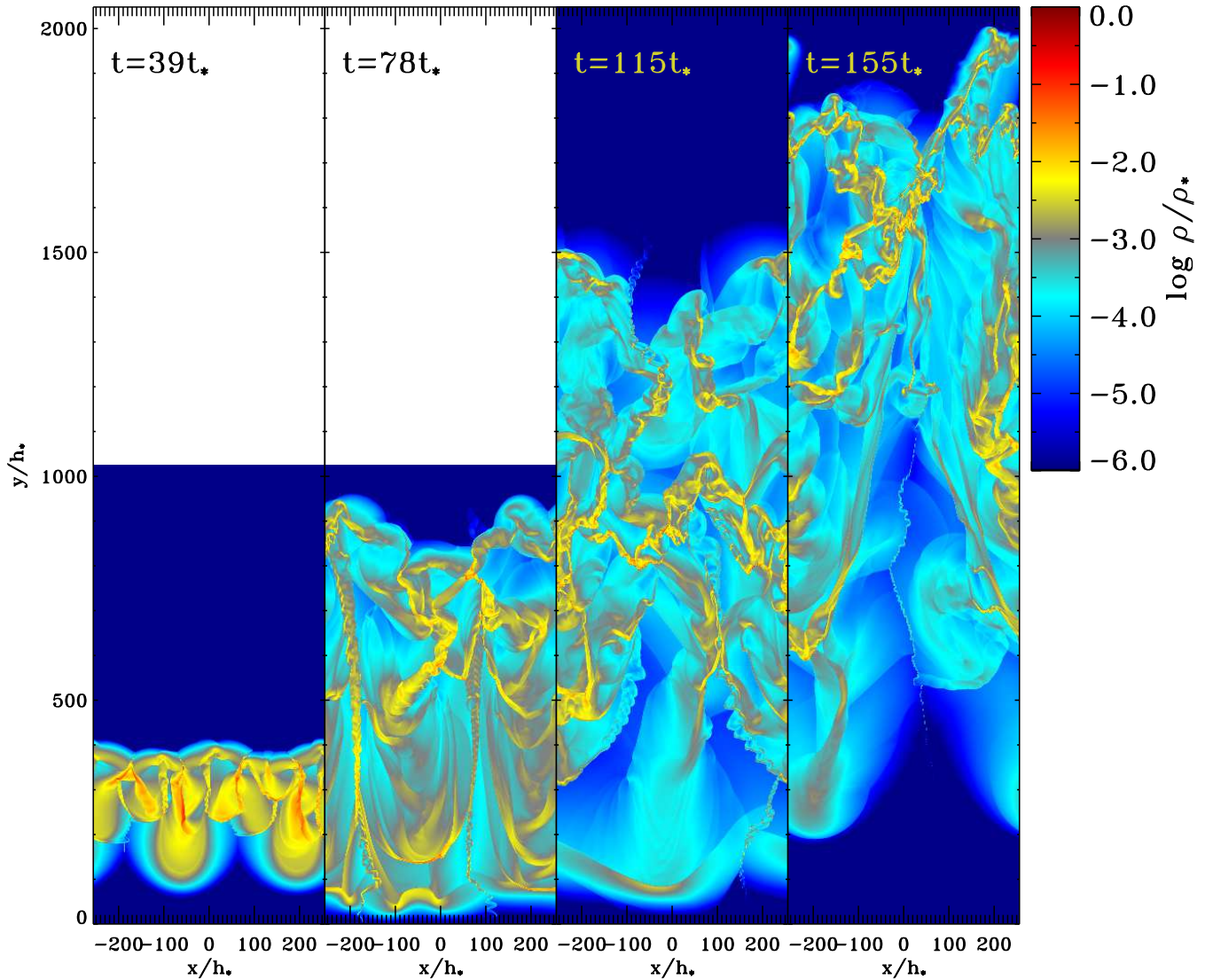}
\caption{Density distribution $\rho$ for four snapshots from the
  T3F0.5VET simulation.  Each panel shows the full simulation domain
  at the labeled times.  Due to gas with high $\rho$ reaching the
  vertical boundary, the simulation was restarted at $t=80 t_*$ in a
  domain with double the vertical extent, as described in the text.
\label{f:denst3f0.5vet}}
\end{figure*}

As in the T10F0.02VET run, there is an initial increase in $T_{\rm rad}$ at
the base of the domain due to the finite optical depth, constant
incoming flux, and close thermal coupling between gas, dust, and
radiation.  This increase in $T$ drives an increase in $\kappa_{\rm
  R}$, resulting in an increase of the radiation force above the
gravitational force.  Hence, the lower part of the atmosphere quickly
becomes super-Eddington, and the vast majority of the mass is driven
upward in a thin shell.

Figure~\ref{f:t3f0.5fld} shows snapshots of $\rho$ and $T_{\rm rad}$
from four different times.  By $25 t_*$, the shell has already begun to
break up, with some material running behind the shell (or even falling
back) in a number of plumes.  This behavior is consistent with
expectations that the shell should be subject to the RTI when
accelerated against gravity by the radiation forces under these
conditions.  (See section \ref{rti} for further discussion.) Our
results are also qualitatively consistent with those of KT12, who also
attribute the non-linear structure to RTI. The growth of non-linear
structure in our run is somewhat faster and the structures are less
uniform, consistent with our additional random perturbations, which
seed the growth of smaller scale structure.

\begin{figure*}
\includegraphics[width=\textwidth]{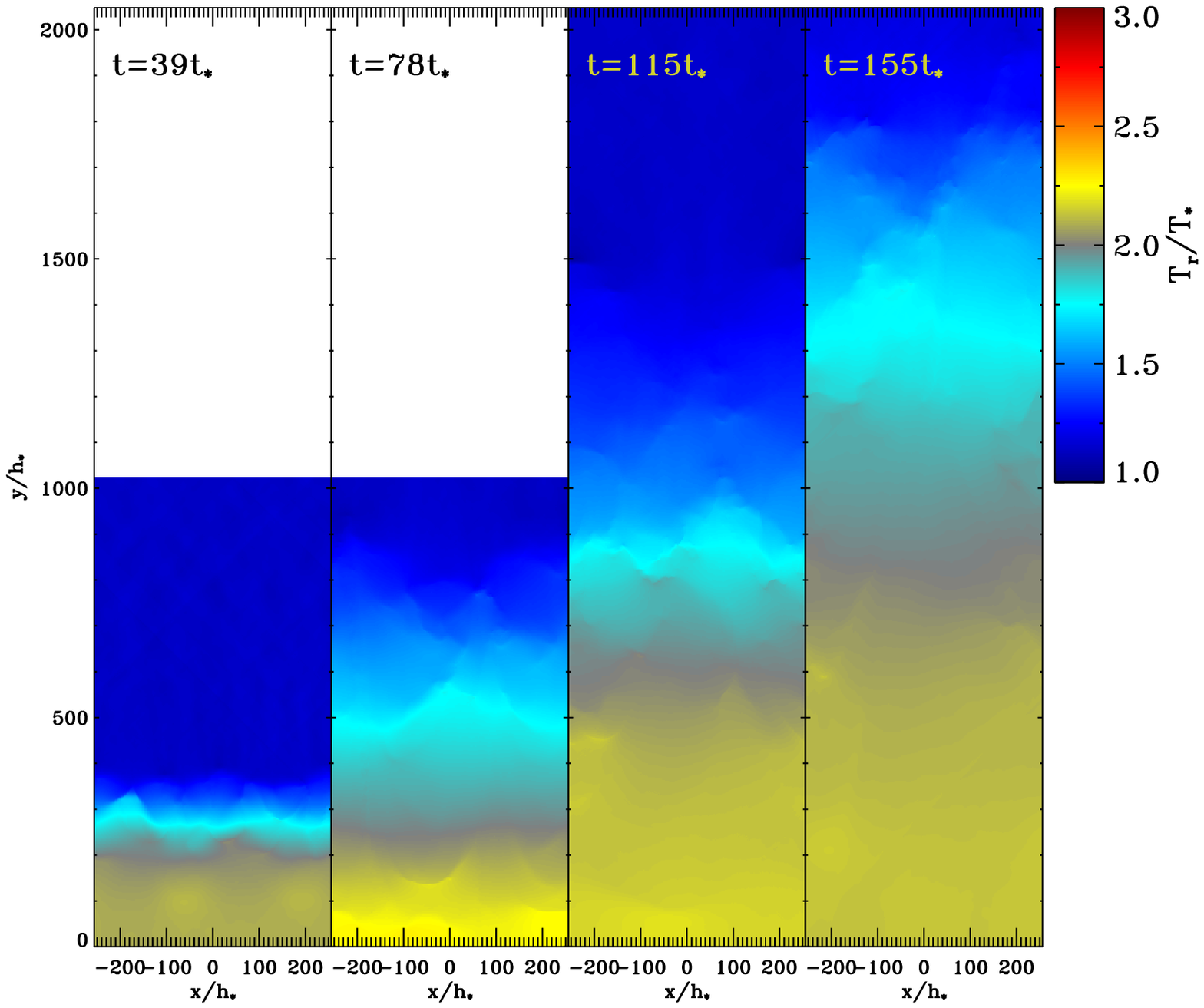}
\caption{Radiation temperature distribution $T_r$ for the four
  snapshots from the T3F0.5VET simulation shown in figure
  \ref{f:denst3f0.5vet}.  Each panel shows the full simulation domain
  at the labeled times.  Due to gas with high $\rho$ reaching the
  vertical boundary, the simulation was restarted at $t=80 t_*$ in a
  domain with double the vertical extent, as described in the text.
\label{f:tradt3f0.5vet}}
\end{figure*}

Subsequent evolution is also consistent with KT12.  The RTI plumes
grow into high density filaments interspersed with lower density
channels.  Since $F_{ry}$ is largest in the low density channels,
there is an anti-correlation of $F_{ry}$ and $\rho$ that reduces the
momentum exchange between the radiation field and gas.  The radiation
forces become sub-Eddington in the optically-thick, high-density
filaments and the majority of the matter falls back by $50 t_*$.  This
behavior is reinforced by a drop in the volume averaged optical depth,
which leads to a drop in $T_{\rm rad}$ (and therefore $T$) at the base
of the atmosphere, further reducing the radiation force.  By $75 t_*$,
almost all of the mass is contained in the bottom $\sim 50 h_*$ of the
domain and remains there for the duration of the run, which we stop at
$200 t_*$.

As expected from the lack of a one-dimensional equilibrium, the matter
at the base of the domain does not settle into a hydrostatic
equilibrium, but instead remains in a turbulent state, with a moderate
velocity dispersion and very little average vertical motion.  Again,
this quasi-steady state turbulent flow is qualitatively and
quantitatively consistent with the results in KT12.

We now turn our attention to the VET run labeled T3F0.5VET.  This
simulation began in a domain with the same dimensions,
resolution, and initial condition as in the T3F0.5FLD run.
Figures~\ref{f:denst3f0.5vet} and \ref{f:tradt3f0.5vet} show four
snapshots of $\rho$ and $T_{\rm rad}$ (respectively) from T3F0.5VET.
Initially, the dynamics of the simulation are qualitatively consistent
with the T3F0.5FLD run.  Again, $T_{\rm rad}$ and $T$ rises rapidly at
the base of the domain and the bulk of the mass is launched upward as
a shell.  The RTI seems to grow slightly slower than in the FLD run,
but is still fairly rapid and significant growth of non-linear
structure is apparent by $39 t_*$.

The difference between the FLD and VET runs become more apparent
in the subsequent evolution.  As in the T3F0.5FLD run, there is 
a flux-density anti-correlation that allows high density filaments
to become sub-Eddington even as the volume averaged structure remains
super-Eddington.  Some of these high-density filaments do fall back
to the base of the domain, where they are reflected by the boundaries.
Others are disrupted and reaccelerated by the radiation field as
they disperse.  The overall evolution leads to a gradual filling
of the volume with $\rho \gtrsim 10^{-4} \rho_*$ gas, although
most of the mass remains in several dense filaments.

At $\sim 83 t_*$, the dense gas reaches the upper boundary of the
initial domain and the assumptions made for the radiation boundary
condition (near vacuum with no incoming radiation) cease to be valid.
This leads to an abrupt, unphysical increase in $T_{\rm rad}$ at the
upper boundary.  Therefore, we restart the simulation at $t=80 t_*$ in
a domain with the $L_y=2048 h_*$ and $N_y=4096$ (i.e.  we double the
height at fixed resolution).  All variables are copied into the bottom
half of the new domain.  Velocities in the upper half of the domain
are set to zero.  All other material variables, $\mathbf{F}_r$, and
$E_r$ are averaged on the upper most grid zones and the upper half of
the domain is uniformly initialized with these values.  The Eddington
tensor is recomputed with the short characteristics module.
Subsequent evolution matches the original evolution until $\sim 83
t_*$ when the breakdown of the vertical boundary condition alters the
evolution in the smaller domain.  We continue our integration in the
larger domain until $158 t_*$, when the dense gas again reaches the
upper boundary of the larger domain.  Again, most of the mass is
concentrated in a few of the densest filaments, but the remainder is
nearly volume filling with $\rho \gtrsim 10^{-4} \rho_*$, except for
transient periods when most of the volume near the bottom of the
domain can have $\rho \lesssim 10^{-6} \rho_*$.

Figure \ref{f:tradt3f0.5vet} shows that there is relatively little
horizontal variation in $T_{\rm rad}$, consistent with T3F0.5FLD and
KT12.  In contrast to T3F0.5FLD, the vertical distribution spreads
out vertically with time as matter fills the domain, but temperature
at the base of the domain remains relatively constant, with $T \simeq
T_{\rm rad} \sim 2.2 T_*$.  Modest, but short-lived increases in
$T_{\rm rad}$ are seen, and correspond to fall back and reflection of
dense filaments at the lower boundary (see e.g. the second panel at
$78 t_*$).  These values are modestly higher than the mean values
of $T_{\rm rad}$ at the base in T3F0.5FLD, although there is greater
fluctuation in $T_{\rm rad}$ in the T3F0.5FLD run.

\begin{figure}[ht]
\includegraphics[width=0.5\textwidth]{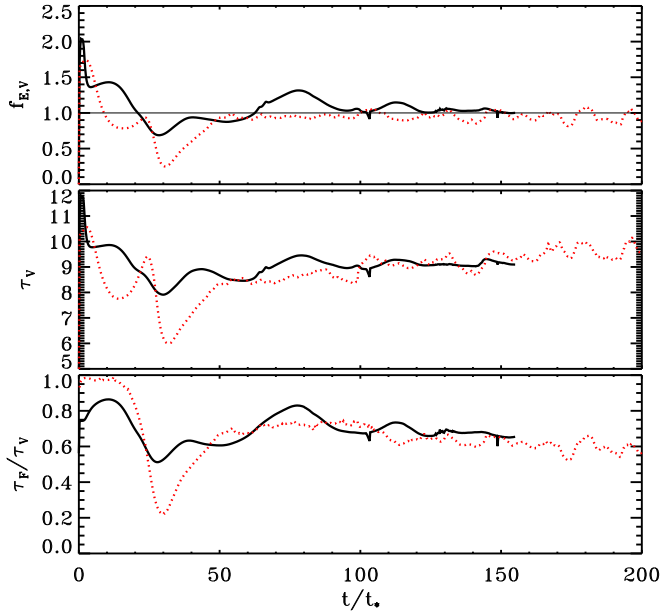}
\caption{Top panel: Volume-averaged Eddington ratio versus time for
  the T3F0.5VET (solid black) and T3F0.5FLD (dotted red) simulation.
  Middle panel: Volume-averaged optical depth versus time for the same
  simulations as in the top panel.  Bottom panel: Ratio of the flux-weighted
  mean optical depth to the volume-averaged optical depth versus time for
  the same
  simulations as in the upper  panels.
\label{f:feddcomp}}
\end{figure}

\begin{figure}
\includegraphics[width=0.5\textwidth]{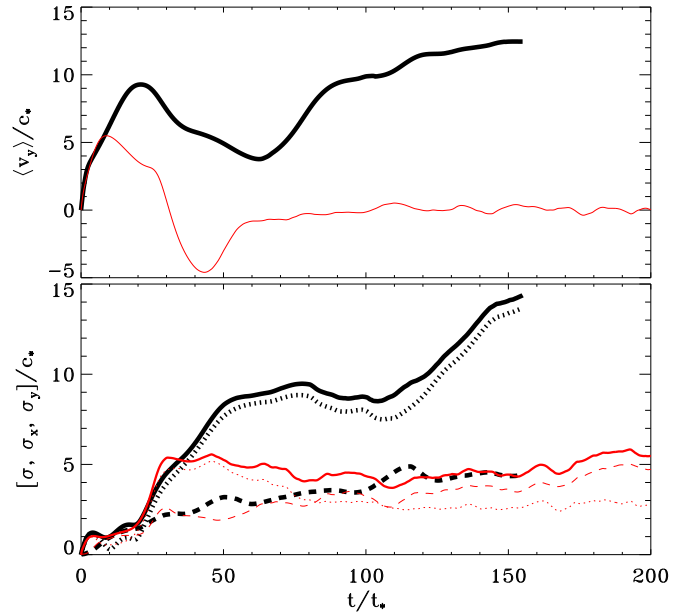}
\caption{Top panel: Mass-weighted mean velocity versus time for the
  for the T3F0.5VET (thick, black curve) and T3F0.5FLD (thin, red
  curve) simulations.  Bottom panel: Mass-weighted velocity dispersion
  versus time for the same runs. For each simulation, the curves
  correspond to $\sigma$ (solid), $\sigma_x$ (dashed), and $\sigma_y$
  (dotted).  In both panels, the velocities are normalized to the
  initial isothermal velocity $c_* =0.54 \; \rm km \; s^{-1}$.
\label{f:sigt3f0.5}}
\end{figure}

Figure~\ref{f:feddcomp} shows a comparison of volume averaged
quantities and their ratios from the T3F0.5FLD and T3F0.5VET
simulations.  The top panel compares the characteristic Eddington
ratio
\begin{equation}
f_{\rm E, V}=\frac{\langle \kappa_{\rm R} \rho F_{ry} \rangle}{c g \langle \rho \rangle}
\end{equation}
where $\langle \; \rangle$ denote volume averages (e.g. $\langle \rho
\rangle = L_x^{-1} L_y^{-1} \int \int \rho \; dx dy$).  Both runs have
an initial super-Eddington period that is followed by a gradual
decline to sub-Eddington values, although this decline is more rapid
and deeper in the FLD case.  Both simulations show a return to an
Eddington ratio near unity, although T3F0.5FLD tends to have $f_{\rm
  E, V}$ fluctuating near unity, while T3F0.5VET tends to remain
moderately super-Eddington for most of the run.  Although the
differences are rather modest, the upshot is that T3F0.5VET receives a
nearly continual, modest upward acceleration while T3F0.5FLD
settles into a quasi equilibrium state.  This evolution is apparent in
the evolution of $\langle v_y \rangle$ in figure~\ref{f:sigt3f0.5}.
The trend in $f_{\rm E, V}$ suggests that T3F0.5VET might settle into
a similar equilibrium, but with a much larger scale height and
higher velocity dispersion.

In order to explore the evolution and impact of optical depth
variations, it is useful to define two characteristic averages
for the optical depth at the base of the domain.  The first is the
volume-weighted mean  optical depth
\begin{equation}
\tau_{\rm V} = L_y \langle \kappa_{\rm R} \rho \rangle,
\end{equation}
and the second is the flux-weighted mean optical depth
\begin{equation}
\tau_{\rm F} = L_y \frac{\langle \kappa_{\rm R} \rho F_{ry}\rangle}
{\langle F_{ry} \rangle}.
\end{equation}

The middle panel of figure~\ref{f:feddcomp} shows that $\tau_{\rm V}$
correlates closely with $f_{\rm E, V}$ as both evolve over the course
of the simulations.  Large values of $\tau_{\rm V}$ correspond to
large values of $\kappa_{\rm R} \propto T^2$, because $T \simeq T_{\rm
  rad}$ and $ T_{\rm rad}$ is roughly proportional to $\tau_{\rm
  V}^{1/4}$ throughout much of the domain.  Note, however, that it is
not quite true that $\tau_{\rm V}$ determines the Eddington ratio as
$f_{\rm E,V}$ can be larger in T3F0.5VET, even when $\tau_{\rm V}$ is
greater for T3F0.5FLD.  This occurs primarily because the nature of
anti-correlation between $F_{\rm ry}$ and $\rho$ differs in the
two runs.

The bottom panel of figure~\ref{f:feddcomp} shows the ratio $\tau_{\rm
  F}/\tau_{\rm V}$.  Note that $\tau_{\rm F}$ multiplies $\langle
F_{ry} \rangle = F_*$ to give the momentum per unit area transfered
from the radiation to the gas while $\tau_{\rm V}$ would be the
characteristic value for the total infrared optical depth $\tau_{\rm IR}$ in
a uniform medium.  Hence, this ratio gives an estimate of how much the
flux -- density anticorrelation reduces the momentum coupling between
radiation and gas.  Our $\tau_{\rm F} = f_{\rm trap} +1$, where
$f_{\rm trap}$ is trapping factor discussed in KT12.  Our estimate of
$\tau_{\rm F} \simeq 6$ agrees with KT12's $f_{\rm trap}=5$ for
$\tau_*=3$, $f_{E,*}=0.5$.

Overall, figure~\ref{f:feddcomp} leaves the impression that the FLD
and VET simulations yield very similar results, particularly at late
times, in contrast with the impression provided by comparison of
figure~\ref{f:t3f0.5fld} to figures~\ref{f:denst3f0.5vet} and
\ref{f:tradt3f0.5vet}.  The key point is that even modest variations
$\mathbf{F}_r$ can lead to rather large differences in the outcome
when systems are near an Eddington ratio of unity.

The differences are somewhat more apparent in
figure~\ref{f:sigt3f0.5}, which shows the evolution of $\langle v_y
\rangle$ and $\sigma$.  For both simulations, the evolution of
$\langle v_y \rangle$ is largely determined by the value of $f_{\rm E,
  V}$ in figure~\ref{f:feddcomp}.  When $f_{\rm E, V} > 1$, $\langle
v_y \rangle$ increases and when $f_{\rm E, V} < 1$, $\langle v_y
\rangle$ decreases.  In both simulations, there is an early period of
transient growth, followed by a decline after the RTI sets in.  For
T3F0.5FLD, $\langle v_y \rangle$ becomes negative as most of the mass
falls back, but eventually settles into a quasi-steady state, with
$\langle v_y \rangle$ simply fluctuating near zero.  For T3F0.5VET,
$\langle v_y \rangle$ grows for the majority of the run, but appears to
be flattening out at late times when $f_{\rm E, V} \sim 1$.

For both runs, the $x$ and $y$ components of $\sigma$ show a modest
initial increase, followed by a faster rises in $\sigma_y$ as the RTI
sets in and the shells start to break apart.  The subsequent evolution
for T3F0.5FLD involves a weak decline in $\sigma_y$ compensated by a
slow increase in $\sigma_x$.  These trends roughly cancel so that
$\sigma$ fluctuates about $\sim 5 c_*$ for the majority of the
simulation, consistent with the results of KT12.  The T3F0.5VET
evolution displays a more continuous rise in $\sigma_y$, with only a
brief drop after $\sim 75 t_*$, when $\langle v_y \rangle$ increases
quickly during a period of moderately high $f_{\rm E, V}$.  We find a
slow, continuous rise in $\sigma_x$, similar to the T3F0.5FLD.  When
we need to stop the run shortly after $150 t_*$, $\sigma$ is still
increasing and already a factor of $\sim 3$ larger than in T3F0.5FLD.

\section{Discussion}
\label{discussion}

\subsection{The Dependence on the Radiation Transfer Method}
\label{comparison}

In this work, we have revisited the calculations of KT12, but have
been unable to reproduce all of their results with our more accurate
VET radiation transfer algorithm.  In contrast, our own implementation
of the FLD algorithm seems to agree well with their calculations,
indicating that discrepancies arise from the use of the FLD
approximation, rather than implementation differences between the
Athena and ORION codes.  Although we reproduce several aspects of
KT12's results (see section~\ref{results}), in this section we focus
primarily on the discrepancies, and describe the deficiencies of the
FLD algorithm that we believe are primarily responsible.

\begin{figure}
\includegraphics[width=0.5\textwidth]{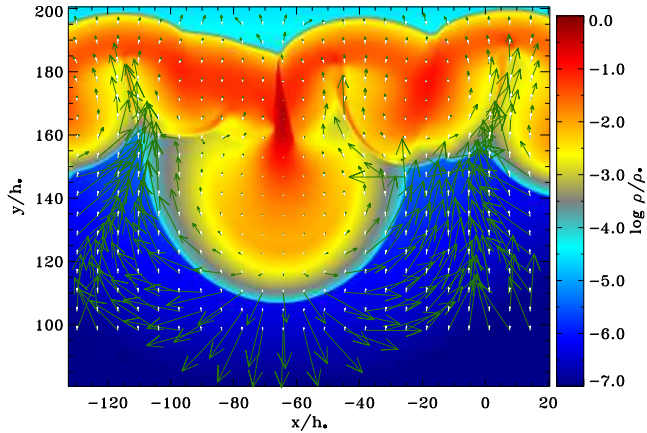}

\caption{ Map of density and radiative flux as functions of position
  for a snapshot from the T3F0.5VET simulation.  Color denotes
  $\rho/\rho_*$ and the two sets of vectors show the direction of
  $\mathbf{F}_r$, with length scaled by magnitude.  The white vectors
  are values of $\mathbf{F}_r$ computed by the VET algorithm and the
  green vectors represent the fluxes that would be inferred from FLD
  (equation [\ref{eq:fld}]) using the $E_r$ in the VET run.  To show
  detail, we only plot a small fraction of the domain.  The snaphsot
  corresponds to $t=25 t_*$, when the disruption of the shell is well
  underway, but not complete.\label{f:dens_flux}}
\end{figure}

First, we note that our VET calculations do closely reproduce the KT12
results in the stable regime.  Therefore, qualitative differences in
the evolution arise from the development of non-linear structure,
driven by the RTI.  Although the development of the RTI is slightly
slower in the VET case, some general aspects of the evolution are
similar to the FLD simulations: a thin shell develops, becomes
unstable to RTI, breaks up into high density filaments interspersed
with low-density channels, and eventually settles to state with
$f_{\rm E,V} \sim 1$ at late times.  

In spite of these qualitative similarities, the differences in
transfer methods have a fairly striking impact on the evolution of the
velocity and spatial distributions of the gas.  For the majority of
the VET simulation, the radiation force exceeds gravity and the net
vertical velocity is always positive.  In contrast, the FLD results
only see a transient acceleration phase followed by fallback of most
of the gas, and settle into quasi-steady state turbulence with a scale
height and velocity dispersion which are much too low to explain
observed systems.

We would like to understand what aspects of the FLD method lead to
these discrepancies in evolution.  We note that since both simulations
reach Eddington ratios near unity, only modest differences are needed
to produce divergent outcomes. Figure \ref{f:dens_flux} shows a
snapshot of $\rho$ for the T3F0.5VET run at $t=25 t_*$.  We focus on a
small fraction of the domain where the shell is being disrupted.  The
white vectors denote the $\mathbf{F}_r$ values, scaled by magnitude,
from the VET calculation.  For comparison, we also plot green arrows
showing what $\mathbf{F}_r$ would be using equation (\ref{eq:fld})
with the values $E_r$ in the VET run.

\begin{figure*}
\includegraphics[width=\textwidth]{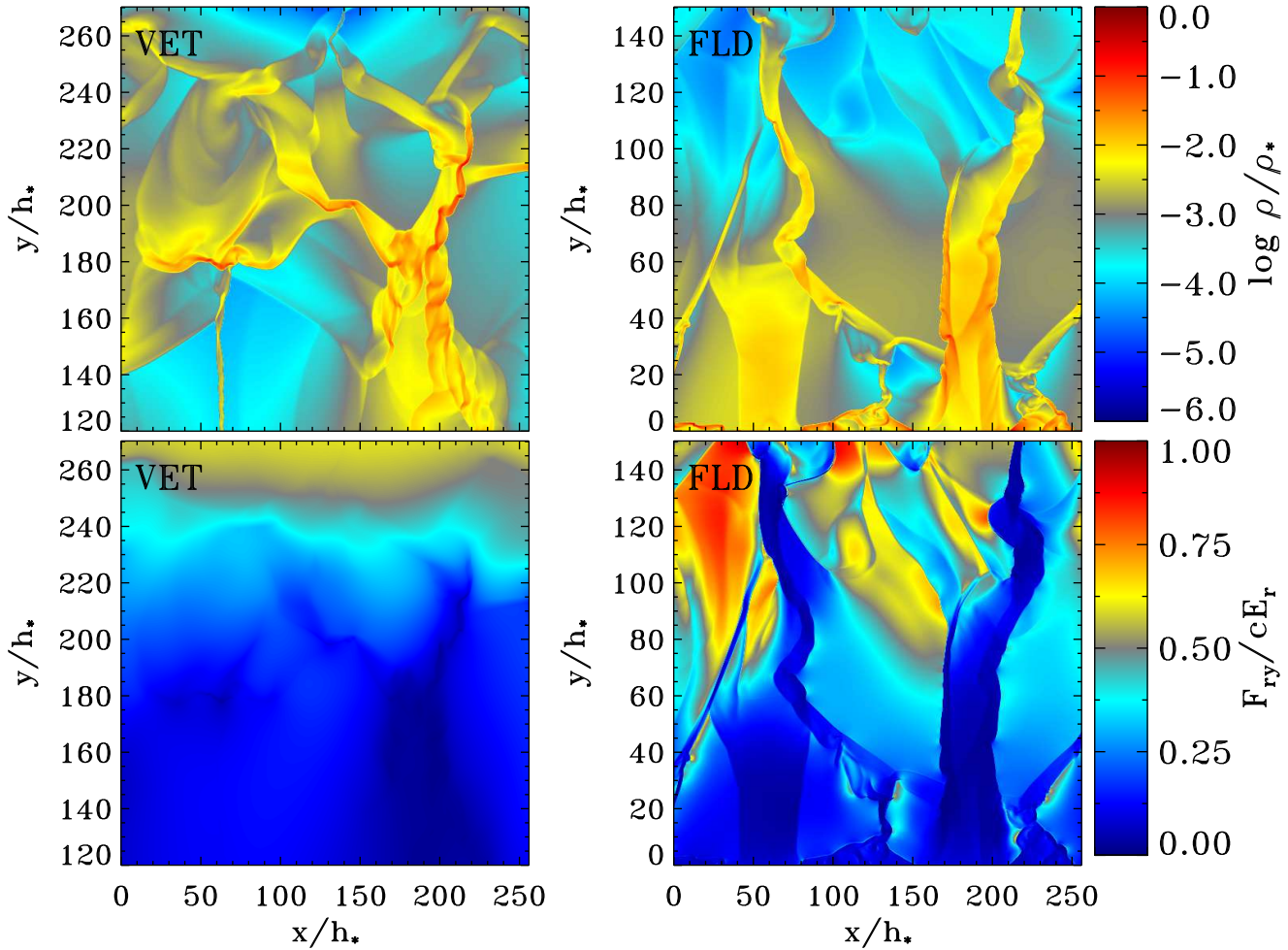}

\caption{
Top panels: Maps of density $\rho$ for snapshots of the T3F0.5VET
(left) and T3F0.5FLD (right) runs.  Bottom panels: Maps of the ratio
of the vertical component of flux to energy density $F_{ry}/cE_r$ for
snapshots of the T3F0.5VET (left) and T3F0.5FLD (right) runs.  The $256
h_* \times 300 h_*$ subsets of the domain are shown at $t=50 t_*$ in
T3F0.5VET run and at $t=42 t_*$ for the T3F0.5FLD.  Note that the
bottom of the T3F0.5FLD subset corresponds to the base of the domain
while the bottom of the T3F0.5VET subset is $120 h_*$ above the base
of the domain.  There is a clearly correlation between $\rho$ and
$F_{ry}/cE_r$ in both calculations, but the degree of correlation is
much stronger in the T3F0.5FLD run.\label{f:limiter}}
\end{figure*}

In the densest regions, our VET calculations are nearly in the
diffusion limit where the FLD method is reliable, and the two sets of
vectors are in approximate agreement.  In lower density regions, where
the diffusion approximation is poor, the magnitude of the FLD flux is
much larger and the directions may be significantly different, even
anti-parallel.  For example, below and adjacent to the large plume in
the center of the figure, the FLD fluxes point downward or horizontal
while the VET fluxes point nearly upward almost everywhere in the
domain.  Since the radiation forces are proportional to
$\mathbf{F}_r$, the FLD fluxes would not be opposing the downward
motion of the plume to the same extent as the VET flux and may even be
reinforcing it.  In contrast, the underdense regions near $x \sim 0,
\; -115 h_*$ and with $160 \le y/h_* \le 190$ have implied FLD fluxes
that significantly exceed the VET fluxes.  This means that (relative
to VET) the FLD radiation forces would be more effective at
reinforcing the development of the low density channels that are
forming.  Overall, the tendency is for FLD to reinforce the
development of non-linear structure to a greater extent than the VET
algorithm, which is consistent with the faster non-linear development
of the RTI in the FLD runs.

One objection to the comparison in figure~\ref{f:dens_flux} is that
the values of $E_r$ in the FLD run will not generally be the same as
those in the VET run.  Figure~\ref{f:limiter} avoids this issue by
comparing $\rho$ and $F_{r y}/E_r c$ in the T3F0.5VET run to the same
quantities in the T3F0.5FLD run.  In both cases, we focus on a $256
h_* \times 300 h_*$ subsets of the simulation domains.  The upper
boundaries of these subsets are chosen to so that less than an optical
depth of intervening matter lies between the top of the subset and the
top boundary of the simulation domain, where vacuum boundary
conditions are imposed.

Since the mean-free-path of photons is proportional to $1/(\kappa
\rho)$, $F_{ry}$ will generally be larger when $\rho$ is smaller and
vice-versa.  Such an anti-correlation is apparent in both runs and is
particularly clear when we scale $F_{ry}$ with $E_r$.  Such
anti-correlations are expected to be particularly strong if the
gradient in $E_r$ is relatively uniform and the diffusion limit
applies (equation [\ref{eq:fld}] with $\lambda \rightarrow 1/3$).
However, the optical depths across most of the filaments seen in
figure~\ref{f:limiter} are of order unity or less, so the diffusion
limit is not valid and the radiation flux is expected to be less
sensitive to local values of $\rho$.  Instead it is determined by the
geometry of sources of strong emission and the integrated optical
depth along different lines-of-site to these sources.  The local value
of the radiation field is determined by non-local properties of the
flow.  The VET algorithm accurately captures the variation of $F_{r
  y}/E_r c$ in this regime, because the Eddington tensor follows from
the calculation of the angle dependent radiation transfer equation.
In contrast, the FLD radiation field is highly constrained by the ad
hoc assumptions that underlie the approximation.  There is a single
preferred direction (parallel to $\nabla E_r$) and the ratio $F_{r
  y}/E_r c$ is determined by $R$ (equation [\ref{eq:limiter}]). Since
$|\nabla E_r|/E_r$ and $T$ are relatively uniform, the sharp variation
of $\rho$ dominates the variation of $R$ and, therefore, $\lambda(R)$.
This stronge dependence of $\lambda(R)$ on $\rho$ leads to an
unphysically high level of anti-correlation between $F_{r y}/E_r c$
and $\rho$ in low-to-moderate optical depth regions.

The upshot is that FLD tends to overestimate the contrast in
$\mathbf{F}_r$ between the high density filaments and the low density
channels, where most of the radiation escapes.  In effect, the FLD
radiation field is more effective at ``punching holes'' through the
$\rho$ distribution and then reinforces the resulting channels to a
greater degree than in the more accurate VET calculations.  This
allows more of the radiation to escape through the low density channels,
leading to a lower average radiation force for the majority of the gas.

Finally, we note that the FLD and VET methods agree well in regions
where structures are moderately optically thick ($\tau \gtrsim 5$) and
the diffusion approximation applies.  This suggests that the FLD
results may still be relevant for systems where the photospheric
Eddington ratio is low.  In these systems (i.e. systems with low $f_*$
and large $\tau_*$), the radiation force can only exceed the
gravitational force at high optical depth, where $T$, and therefore
$\kappa_{\rm R}$, is sufficiently large.

\subsection{Implications for Observed Systems and Subgrid Models of
Radiation Feedback}

Since our VET solutions differ in important respects from earlier
results using FLD, it is important to reexamine the implications of
RTI for radiation feedback.  Although the current simulation setup is
conducive to exploring the role of RTI, the choice to start with a
hydrostatic initial condition limits the relevance to observations of
real star forming galaxies.  This is because the hydrostatic length
scale in the problem is $h_*=6.4 \times 10^{-4} \rm pc$ (for $T_*=82
K$ and $f_{\rm E,*}=0.5$ ) and $L_z = 2048 h_* = 1.3 \rm pc$.  Hence
the domain is nearly two orders of magnitude too small to contain a
realistic ULIRG disk.  Other simplifications, such as the assumption
of a constant $g$, approximate grey opacity, and the constant incident
infrared flux may also have a strong effect on the evolution.

Nevertheless, we can reach some tentative conclusions.  One quantity
of interest is the mass-weighted velocity dispersion $\sigma$.  Since
$\sigma$ is growing throughout most of the simulation and $\langle v_y
\rangle \gtrsim \sigma$ at the simulations's end, we view our maximum
$\sigma$ as a lower bound rather than a characteristic estimate.  We
find $\sigma \simeq 14 c_* \simeq 7.5 \; \rm km \; s^{-1}$.  This is
about an order-of-magnitude lower than the values typical inferred
\citep[e.g.][]{DownesSolomon1998}, although this is unsurprising given
the small size of the simulation domain.  It is difficult to assess
what implications the trend towards $f_{\rm E,V} \sim 1$ will have for
the subsequent evolution of $\langle v_y \rangle$ and $\sigma$.  The
system is very sensitive to the precise value of $f_{\rm E,V}$ when
$g$ is constant.  If (on average) $f_{\rm E,V}$ continues to be
slightly greater than one, then acceleration will continue and we
would expect the scale height and velocities to grow with time
indefinitely.  In a real system, the vertical variation of $g$ would
presumably play a significant role in determining an equilibrium scale
height and velocity dispersion.

A second question of interest is what is the correct efficiency (or
``trapping'') factor to use in assessing the momentum imparted by
radiation pressure.  In feedback models for optically thick
environments, the rate of momentum injection has been parameterized as
\citep[see e.g][and references therein]{Hopkinsetal2011}
\begin{equation}
\frac{d p}{d t} = (1 + \eta \tau_{\rm IR}) \frac{L}{c},\label{eq:feedback}
\end{equation}
where $dp/dt$ is an estimate of the momentum transferred to the
interstellar gas, $\tau_{\rm IR}$ is an estimate of the integrated
optical depth through the dusty gas at infrared wavelengths, $L$ is
luminosity of a star (or star cluster), and $\eta$ is an ad hoc
reduction factor ($\eta \lesssim 1$) included to account for
inhomogeneities in the surrounding gas.  Since $\tau_{\rm V}$
corresponds to a volume-weighted average total optical depth, one can
estimate $\tau_{\rm IR} \simeq \tau_{\rm V}$, while $\tau_{\rm F}$
represents the effective optical depth for momentum exchange.
Therefore, we estimate the $\eta$ in equation (\ref{eq:feedback}) as
the ratio $\tau_{\rm F}/\tau_{\rm V} = 0.68$, where we have time
averaged the VET run for $t > 100 t_*$.

A value of $\eta \simeq 0.68$ is in the same range that was considered
by \citet{Hopkinsetal2011} in their calculations.  As noted in
section~\ref{stable}, our value of $\tau_{\rm F} \simeq 6$ agrees with
KT12's $f_{\rm trap}=5$ (since $\tau_{\rm F} = f_{\rm trap} +1$).
However, in their discussion KT12 argue that \citet{Hopkinsetal2011}
and others likely overestimate $\eta$.  The origin of this apparent
discrepancy lies partly in how one estimates $\tau_{\rm IR}$ and
partly in the dependence of $\tau_{\rm V}$ and $\tau_{\rm F}$ on
$\tau_*$ and $f_{\rm E,*}$.  First, KT12 estimate $\tau_{\rm IR}
\approx \kappa(T_{\rm mp}) \Sigma$, where $T_{\rm mp}$ is the
temperature at the base of the simulation domain.  For our simulation,
this quantity is a factor of $\sim 1.5$ larger than $\tau_{\rm V}$ for
T3F0.5VET.

KT12 also perform a number of calculations with $\tau_*=10$, which are
associated with lower values of $\eta$.  Although we consider only a
single parameter set with $\tau_*=3$, we can derive approximate
scalings for $\eta$ with $\tau_*$ and $f_{E,*}$ by assuming that all
simulations will approach a characteristic horizontally average
vertical profile.  At late times our T3F0.5VET run can be fit
approximately by\footnote{Note that equation (\ref{eq:temp2}) implies
  $E_r \simeq 2 F_{ry}/c$, which is roughly the ratio returned by our
  short-characteristics calculation at the top of the domain.  For an
  approximately isotropically emitting, semi-infinite atmosphere,
  which is essentially what our periodic horizontal boundary
  conditions provide, this agrees well with standard grey treatments
  \citep{Mihalas1978}.  This is greater than the $E_r \simeq F_{ry}/c$
  assumed in FLD calculations, a limit that is only obtained at large
  distances from a point source.}
\begin{equation}
\overline{T}^4=T_*^4 3 \left(\frac{2}{3} +\overline{\tau}_{\rm F}  \right)\label{eq:temp2},
\end{equation}
where quantities with an overbar correspond to $y$-dependent,
horizontally average quantities (e.g. $\overline{\rho} = L_x^{-1} \int
\rho \; dx$).  The exceptions are
\begin{eqnarray}
\overline{\tau}_{\rm F} & \equiv & \int \frac{\overline{\kappa_{\rm R} \rho F_{ry}}}
{\overline{F_{ry}}} \; dz, \nonumber \\
\overline{\tau} & \equiv &\int \overline{\kappa_{\rm R} \rho} \; dz.
\end{eqnarray}

To a reasonable approximation, we find that
\begin{equation}
\overline{\tau}_{\rm F}\simeq\left\{\begin{array}{ll} \overline{\tau};
    &  \overline{\tau} < 1 \\
\eta \overline{\tau}
      & \overline{\tau} \gtrsim 1,
\end{array}\right.
\end{equation}
with a constant $\eta \simeq \tau_{\rm F}/\tau_{\rm V}$.  Using the fact that
\begin{equation}
\frac{\partial \overline{\tau}}{\partial m} = \kappa_{\rm R,*} 
\left(\frac{\overline{T}}{T_*}\right)^2,
\end{equation}
with $d m = - \overline{\rho} dz$, we find
\begin{equation}
\overline{\tau} = \frac{3}{4}\eta \left(\kappa_{\rm R,*} m\right)^2
+\sqrt{2} \kappa_{\rm R,*} m.\label{eq:taubar}
\end{equation}
At the base of the domain, $m=\tau_*/\kappa_{\rm R,*}$ and
$\overline{\tau}(0) = \frac{3}{4}\eta \tau_*^2 +\sqrt{2} \tau_* \simeq
8.8$ for $\tau_*=3$, which agrees with our measurement of $\tau_{\rm V}
\simeq 9$ in T3F0.5VET.

If we assume that all of our simulations will approach
$f_{\rm E,V} \simeq 1$ at late times we can infer that
\begin{equation}
\tau_{\rm F} \approx \frac{\tau_*}{f_{\rm E,*}}=\frac{c g}{F_*} \Sigma,\label{eq:tauf}
\end{equation}
which is equivalent to KT12's equation (43) with $\langle f_E \rangle
=1$.  Thus, for a given input flux $F_*$ and constant gravitational
acceleration $g$, we can define a characteristic opacity $\kappa_{\rm
  E}=c g/F_*$, for which radiative acceleration and gravitational
acceleration balance.  Then the optical depth $\tau_{\rm F}
=\kappa_{\rm E} \Sigma$ can be used in place of $\eta \tau_{\rm IR}$
in equation (\ref{eq:feedback}).  In our calculation $\kappa_{\rm R}
\sim \kappa_{\rm E}$ near the photosphere where $\overline{\tau} \sim
1$.  This is in approximate agreement with conclusion of
\citet{KrumholzThompson2013} that the photospheric opacity is the
relevant opacity for evaluating momentum feedback, although it is not
clear how precisely this would generalize to other parameters.

Alternatively, we can provide an estimate for $\eta$ by combing
equations (\ref{eq:taubar}) and (\ref{eq:tauf}) to obtain
\begin{equation}
\eta \approx \frac{2}{3\tau_*}\left(\sqrt{2+3\tau_*/f_{\rm E,*}}-\sqrt{2} \right),
\end{equation}
which implies $\eta =0.68$ for $\tau_*=3$ and $f_{\rm E,*}=0.5$, in
good agreement with the simulation results.  Hence, $\eta$ would be a
function of $\tau_*$ and $f_{\rm E,*}$, with $\eta \propto
\tau_*^{-1/2}$ for $\tau_* \gg 1$, although it should be kept in mind
that $\kappa_{\rm R} \propto T^2$ approximation becomes poor for $T
\gtrsim 150 \rm K$, so there is a limited range where this relation
could be physically relevant. Finally, we can use equations
(\ref{eq:temp2}) and (\ref{eq:tauf}) to estimate the optical depth
using the temperature at the base of the domain
\begin{equation}
\kappa_{\rm R,0} \Sigma = \tau_* \left(2 + 3\frac{\tau_*}{f_{\rm E,*}} \right)^{1/2}.
\end{equation}
This implies $\kappa_{\rm R,0} \Sigma = 13.4$ for $\tau_*=3$ and
$f_{\rm E,*}=0.5$, nearly matching the measured value of $\kappa_{\rm
  R,0} \Sigma =13.6$ from T3F0.5VET.

Another question of interest is whether these simulations are
consistent with the radiation driving of large scale outflows.  Our
results show that the simulations approach an Eddington ratio near
unity for a constant value of $g$.  However, if we assume that the gas
in real ULIRGs is distributed in a disk-like geometry the vertical
component of gravity $g(y)$ will increase from near zero at the
midplane (where $y=0$) to a maximum value set by the overall potential
of the galaxy. It seems plausible that an Eddington ratio near unity
will also be reached in this case, but it is not clear what would pick
out the specific value of $g_0$ where the radiation forces and gravity
balance.  However, if this hypothetical equilibrium behaves like our
simulations, we would expect a fraction of the gas to be accelerated
well beyond this characteristic $g_0$, because the opacity typically
increases while $g(y)$ decreases as we approach the midplane.
Therefore, gas in low density channels can be efficiently accelerated
to high velocities, conceivably approaching the escape velocity from
the galaxy.  Such acceleration of low density gas to high velocities
is seen in our simulations.  For example, when we stop the T3F0.5VET
run, 4.1\% and 1.2\% of the gas had $v_y$ greater than $3 \sigma_y$
and $4 \sigma_y$ (respectively).  In typical ULIRGs, only a few
percent of the gas is observed in the cold neutral outflows, so even a
small tail of high velocity gas could explain the observed outflow
rates and velocities if $\sigma_y \sim 100 \; \rm km \; s^{-1}$ can be
obtained in more realistic simulations.

In this respect, a prominent role of RTI in the dynamics could help
solve a potential problem with radiation driving -- that it cannot
simultaneously explain both the presence of a radiation supported,
quasi-equilibrium gas disk and the launching of outflows.  If all the
gas experienced the same radiative acceleration, it would all be in
state of hydrostatic equilibrium or it would all be accelerated in an
outflow.  In an RTI dominated picture, the bulk of the gas can be in a
quasi-equilibrium turbulent state with $\langle v_y \rangle \sim 0$,
but the presence of low density channels with larger than average
$F_{ry}$ could still allow a modest fraction of the gas to be
accelerated out the gas disk.  In principle, this could allow
radiation to launch outflows even in galaxies that are well below
their global Eddington limit \citep[c.f.][]{SocratesSironi2013},
although more realistic calculations are required to assess the
effectiveness of this mechanism.

Finally, we note that all of our analysis and discussion assumes that
only radiation and gravitational forces play a role.  In real systems,
mechanical feedback from stellar winds, supernovae, cosmic rays,
magnetic fields, and other sources may be present. Even if they are
not the dominant mechanisms of momentum transfer to the gas, they may
still have a measurable impact on the gas velocity and density
distributions.  The implication that RTI will generically produce
density inhomogeneities that drive $f_{\rm E,V}$ towards a value $\sim
1$ may break down if any of these other mechanisms are strong enough
to reorient the low density channels and dense filaments.  It seems
much more likely that such reorientation will interfere with, rather
than enhance, the escape of photons, leading to increased coupling and
conceivably to generically super-Eddington configurations.  This
hypothesis can be tested with future calculations.

\subsection{The Role of the Rayleigh-Taylor Instability}
\label{rti}

We attribute the growth of structure in our simulations primarily to
the RTI, but other instability mechanisms may play a role, in
principle.  Section 5.1 of KT12 provides a fairly comprehensive and
persuasive discussion why the RTI is dominant and other instability
mechanisms, such as radiative driving of unstable acoustic modes
\citep{Shaviv2001,BlaesSocrates2003}, are likely to be absent or
unimportant.  Here we simply summarize some of the most salient
results and refer the reader to KT12 for further details.

Although a general solution for linear growth of the radiative RTI has
never been formulated \citep[see
  e.g.][]{MathewsBlumenthal1977,Krolik1977,JacquetKrumholz2011,Jiangetal2013},
numerical experiments \citep{Jiangetal2013} suggest that the
instability will grow at a rate that can be approximated by $t_{\rm
  g}^{-1} \sim \sqrt{2\pi g /\lambda_{\rm x}}$, if $\lambda_{\rm x}$
is a characteristic horizontal wavelength.  This is the analytic
prediction in the optically thick limit of an incompressible flow if
the density contrast at the base of the shell is large (i.e. Atwood
number of unity).  Radiative diffusion effects can slow the rate of
growth, but typically only by factors of order unity unless radiation
pressure is much greater that gas pressure.  For our simulations, this
timescale evaluates to $t_{\rm g} = t_* \sqrt{\lambda_{\rm x}/2\pi
  h_*}$ and we have $t_{\rm g} \sim 6 t_*$ for $\lambda_{\rm x} \sim
256 h_*$.  The growth of the instability on small scales is even
faster and there is ample time for growth of the RTI to explain the
large scale non-linear structure observed at $t=25 t_*$ (and earlier).

\begin{figure*}[ht]
\centering
 \begin{tabular}{cc}
\includegraphics[width=0.5\textwidth]{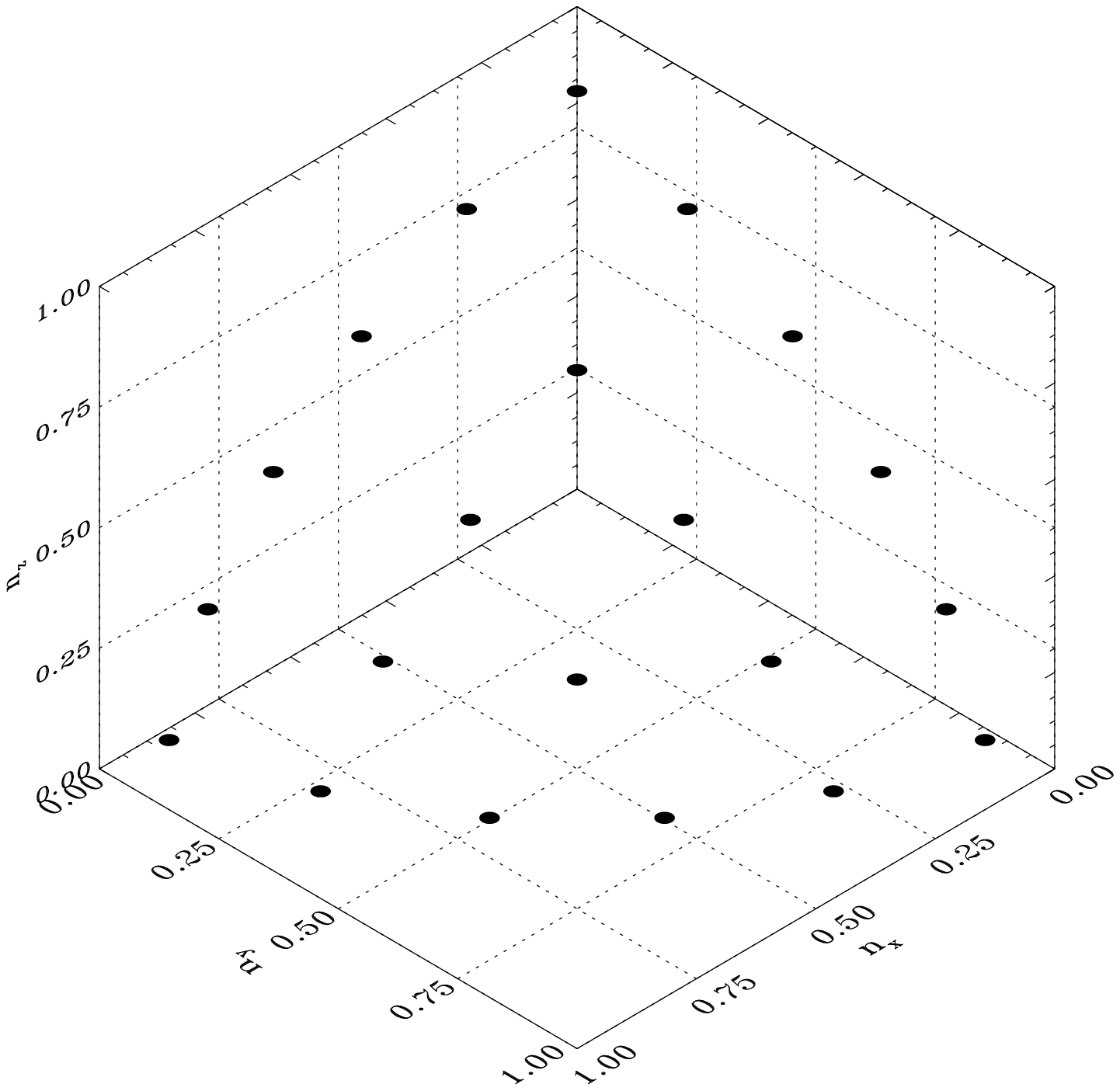} &
\includegraphics[width=0.5\textwidth]{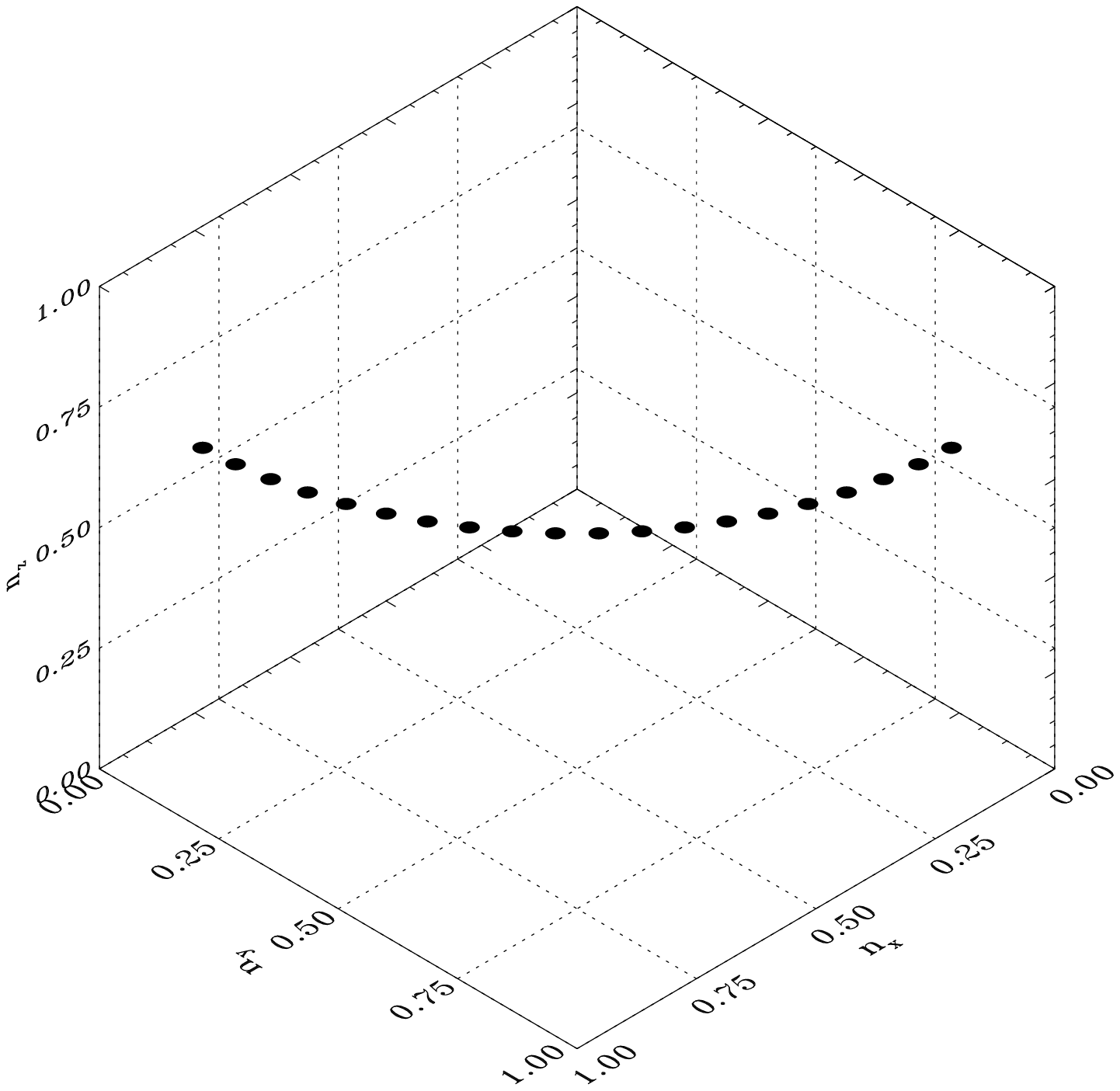}\\
\end{tabular}
\caption{Ray distributions for a single octant of the unit sphere.
  The left panel shows our default distribution used for runs
  T10F0.02VET and T3F0.5VET.  Poloidal angles correspond to abscissas
  of Gauss-Legendre quadrature and azimuthal angles are distributed
  uniformly over the interval 0 to $\pi/2$ within each poloidal
  level. The right panel shows an alternate distribution with a single
  poloidal level ($n_z=1/\sqrt{3}$), which we use to increase the
  effective angular resolution in the $x$--$y$ plane.  The axes
  correspond to the projection of the unit vector corresponding to the
  ray onto the Cartesian axes of the domain e.g. $n_x = \hat{n} \cdot
  \hat{x}$.\label{f:angdist}}
\end{figure*}

Further evidence that the evolution is due to RTI is provided by
calculations we performed with constant, temperature-independent
opacities in the limits of pure scattering or pure absorption.  The
simulation setup for these runs were identical to our T3F0.5VET
calculations except that $f_*=1.25$.  In the standard runs,
with $\kappa_{R} \propto T^2$, radiation forces are largest near the
base of the domain and decrease with height.  Therefore, gas closer to
the base of the domain quickly reaches higher velocities than the
overlying material and leads to the formation of dense shells with
sharp density inversions.  In contrast, the gas in the constant
opacity simulations receives an acceleration that is significantly
more uniform, and no significant density inversions form until very
late in the run.  As a result, there is very little RTI, and the small
scale random perturbations are smoothed out by diffusion before they
can grow appreciably.  Only the largest scales show non-linear
development from the initial sinusoidal perturbation, and the timescale
for this growth is much longer.  Thus, there is strong evidence that
density inversions driven by the $\kappa_{\rm R} \propto T^2$ opacity
law are essential to the dynamics, providing a clear indication that
the RTI is the dominant instability mechanism.

\section{Summary and Conclusions}
\label{summary}

We have considered the role of the Rayleigh-Taylor instability in the
interaction of infrared radiation fields, dust, and gas in rapidly
star-forming environments.  We have focused on the regime of radiation
supported, dense gas that may be present in some systems, such as
ULIRGs.  Our primary results stem from the numerical simulation of
such environments, which are studied in a simplified problem setup
with a constant gravitational acceleration, a constant incident
infrared flux on the base of the domain, and initialized with a
perturbed isothermal atmosphere to match previous calculations
in KT12.

In the stable regime, we find that the atmosphere settles down into an
equilibrium solution in agreement with the previous results.
In the unstable regime, we confirm that the RTI develops and has a
significant impact on subsequent evolution.  However, we find that
after the growth of the RTI, the evolution depends
significantly on the choice of algorithm for modeling radiation transfer.
Our VET simulations show a stronger coupling between radiation and
dusty gas, leading to continuous net upward acceleration of the gas.
No steady state is reached before the end of the calculation, when
high density material had reached the top of the domain.  The mean
velocities and velocity dispersion are both increasing at the end of
the run.  In contrast, our FLD calculations broadly reproduce the
FLD results of KT12, finding weaker coupling between gas and radiation.
This leads to a short initial burst of acceleration, followed by a
period of fallback, finally settling into quasi-steady state
turbulence with zero mean velocity and low velocity dispersion.
As a result, our VET calculations imply a much larger scale height
 and higher velocity dispersion than the FLD-based calculations.

We argue that these discrepancies result from limitations in the
diffusion-based FLD algorithm, which lead to inaccuracies in modeling
how the radiation field responds to structure in the gas distribution
that is a few optical depths or smaller in size.  These errors are
related to the FLD radiation field's tendency to diffuse around
denser, optically-thicker structures even when the diffusion limit
does not apply. Relative to our VET calculations, this leads the FLD
radiation forces to be more effective at opening and reinforcing low
density channels but less effective at disrupting high density
filaments, ultimately reducing the coupling between radiation and
dusty gas.

Despite the discrepancies in the scale height and velocity
distributions, it appears that both the VET and FLD simulations trend
towards an approximate balance between the volume-averaged radiation
and gravitational forces at late times.  If this behavior is general,
it confirms one of the key results of KT12 \citep[see
  also][]{KrumholzThompson2013}, suggesting the rate of momentum
transfer between radiation and dusty gas may scale approximately as
$\sim \tau_{\rm E} L/c$, where $L$ is the luminosity, $\tau_{\rm E}=
\kappa_{\rm E} \Sigma$, $\Sigma$ is the mass surface density, and
$\kappa_{\rm E} = c g/F_*$ is the opacity for which the radiation and
gravitational accelerations balance.  Since $\kappa_{\rm R}$ is
generally larger nearer to the midplane, $\tau_{\rm E}$ will be lower
than estimates of $\tau_{\rm IR}$ that assume volume-average or
mid-plane opacities.  For example, in the simulation parameters
considered here ($\Sigma =1.4 \rm \; g \; cm^{-2}$), we infer a total
optical depth $\tau_{\rm IR} \sim 9$ using a volume-averaged opacity,
which must be reduced by an efficiency factor $\eta \simeq 0.68$ to
obtain the correct rate of momentum transfer.  However, if the
Eddington ratio is always near unity, this efficiency factor may be
lower for gas disks or clouds with larger surface densities.

Although the simulation setup considered here is useful for studying
the development and saturation of the RTI, it is not optimally suited
for making observational predictions.  Future work with more realistic
assumptions, such as a vertically varying gravitational acceleration,
non-grey opacity, distributed radiation sources, and physically
relevant simulation volumes will be necessary to provide robust
predictions and facilitate direct comparison with observations.

\acknowledgements{We thank C.-A. Faucher-Giguer, P. Hopkins,
  C. Matzner, E. Ostriker, and T. Thompson for useful
  discussions. N.M. is supported in part by the Canada Research Chair
  program. Y.F.J. is supported by NASA through Einstein Postdoctoral
  Fellowship grant number PF-140109 awarded by the Chandra X-ray
  Center, which is operated by the Smithsonian Astrophysical
  Observatory for NASA under contract NAS8-03060. Computations were
  performed on the GPC supercomputer at the SciNet HPC
  Consortium. SciNet is funded by the Canada Foundation for Innovation
  under the auspices of Compute Canada, the Government of Ontario,
  Ontario Research Fund-Research Excellence, and the University of
  Toronto. }

\begin{figure}[ht]
\includegraphics[width=0.5\textwidth]{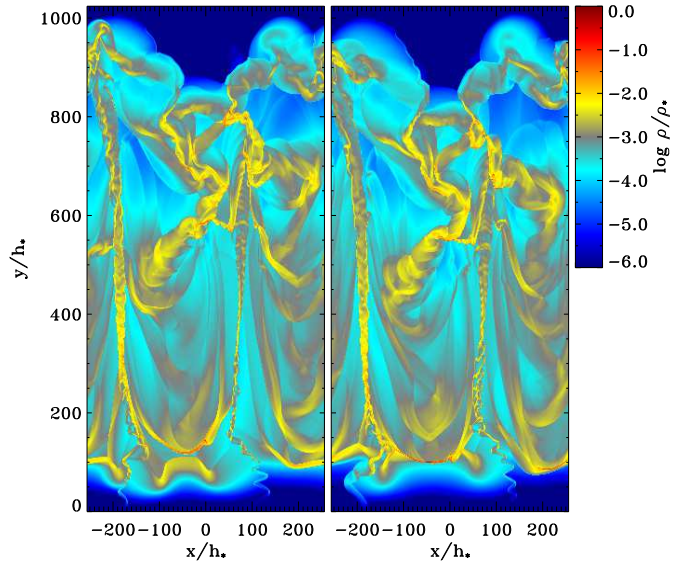}
\caption{Density distribution $\rho$ for snapshots from two versions
  of the T3F0.5VET run shown at $t=80 t_*$.  The left panel shows
  results from the standard run with rays distributed in six poloidal
  levels while the right panel shows results using a single poloidal
  level, as shown in the left and right panels (respectively) of
  figure~\ref{f:angdist}.  As described in section \ref{angquad}, the
  right panel has higher effective angular resolution, reducing
  ray-effects.  The close agreement between the two runs suggests that
  the dynamics are not strongly sensitive to our choice of angular
  grid.
\label{f:anglecomp}}
\end{figure}

\appendix

\subsection{Impact of Angular Resolution}
\label{angquad}

The discussion in section~\ref{comparison} assumes that the VET
radiative transfer algorithm provides a more accurate realization of
the radiation field than the FLD algorithm.  Since we solve the
momentum equation (\ref{eq:radmomentum}), this should be true, as long
as our method for computing the Eddington tensor is valid.  As
discussed in section \ref{shortchar}, the main approximation made in
our computation of the Eddington tensor is the neglect of order $v/c$
terms (which are retained in the radiation moment equations).  We
confirm a posteriori that this is a very good approximation.

The only significant potential problem for our VET calculation would
be if our angular grid of rays under-resolves the radiation field.
Our experience has been that angular resolution used here is more than
adequate for previous problems, but angular resolution requirements
can vary from problem to problem.  In the short characteristics
method, the most apparent features in under-resolved systems are
usually referred to as ``ray-effects'', which correspond to unphysical
anisotropies in the radiation field correlated with the angular grid.
Ray-effects can be particularly problematic when bright point sources
dominate the radiation field and scattering opacity is negligible
\citep[see e.g.][]{LarsenWollaber2008,Finlatoretal2009}.  The
intensity can be overestimated in grid zones that lie along ray
directions that point directly back to the point source and
underestimated in regions between rays, with a relative error that
increases in magnitude with the number of grid zones traversed.  This
leads to the appearance of `rays' or `spokes' in radiation variables
emanating radially outward from the point source.

Scattering introduces angular diffusion that can mitigate ray-effects
\citep[see e.g.][]{LarsenWollaber2008}, but is absent in the current
calculations.  Although bright point sources are also absent, dense
filaments can lead to rather sharp variations in the emissivity.
Ray-effects are most clearly apparent in the low density regions well
above the photosphere, reflecting variations in the emissivity near
the photosphere that propagate along rays in the angular grid into the
regions above.  However, there is very little mass in this region and
radiation is less important because the radiation forces are
sub-Eddington here.

Ray-effects are less apparent below the photosphere, but must be
present at some level.  In order to assess their impact (if any), we
reran the T3F0.5VET run with an alternative ray distribution that has
effectively higher resolution in the $x$--$y$ plane.  Our default ray
distribution is shown in the left panel of figure~\ref{f:angdist}.  It
attempts to cover the unit sphere uniformly, treating all three
spatial directions, including the implied third dimension ($z$) on
equal footing.  This is formally required even though the domain is
two dimensional, because rays traveling at different angles relative
to $\hat{z}$ will travel different total path lengths for a given
displacement in the $x$--$y$ plane.  This means that many of the rays
nearly overlap when projected onto the $x$--$y$ plane, and the
effective angular resolution in the $x$--$y$ plane is closer to $\sim
24$ rays per $2\pi$ radians, even though we have 84 rays total.  The
right panel shows our alternative distribution, in which all rays have
the same projection onto the $z$-axis ($n_z=1/\sqrt{3}$ so that
$f_{zz}=1/3$).  We distribute 80 total rays uniformly in azimuthal
angle $\phi$ ($\tan \phi = n_y/n_x$).  This amounts to a factor $\sim
3$ increase in the effective angular resolution in the $x$--$y$ plane
even though the total number of angles is nearly identical.

The higher effective azimuthal angular resolution significantly reduced
ray-effects above the photosphere, but had relatively little effect on
the Eddington tensor below the photosphere, suggesting ray effects
were already minor in the original run.  Overall, the global properties
and evolution of the two simulation were nearly identical.  A sense of
how similar the runs are is provide by figure~\ref{f:anglecomp}, which
compares snapshots of $\rho$ from the two simulation at $t=80 t_*$.
We therefore conclude that our results are well-converged in terms of
the effective angular resolution of our VET calculations.

\end{CJK*}


\begin{thebibliography}{40}
\expandafter\ifx\csname natexlab\endcsname\relax\def\natexlab#1{#1}\fi

\bibitem[{{Andrews} \& {Thompson}(2011)}]{AndrewsThompson2011}
{Andrews}, B.~H., \& {Thompson}, T.~A. 2011, \apj, 727, 97

\bibitem[{{Blaes} \& {Socrates}(2003)}]{BlaesSocrates2003}
{Blaes}, O., \& {Socrates}, A. 2003, \apj, 596, 509

\bibitem[{{Chandrasekhar}(1961)}]{Chandrasekhar1961}
{Chandrasekhar}, S. 1961, {Hydrodynamic and hydromagnetic stability}

\bibitem[{{Davis} {et~al.}(2012){Davis}, {Stone}, \& {Jiang}}]{Davisetal2012}
{Davis}, S.~W., {Stone}, J.~M., \& {Jiang}, Y.-F. 2012, \apj

\bibitem[{{Downes} \& {Solomon}(1998)}]{DownesSolomon1998}
{Downes}, D., \& {Solomon}, P.~M. 1998, \apj, 507, 615

\bibitem[{{Finlator} {et~al.}(2009){Finlator}, {{\"O}zel}, \&
  {Dav{\'e}}}]{Finlatoretal2009}
{Finlator}, K., {{\"O}zel}, F., \& {Dav{\'e}}, R. 2009, \mnras, 393, 1090

\bibitem[{{Hayes} \& {Norman}(2003)}]{HayesNorman2003}
{Hayes}, J.~C., \& {Norman}, M.~L. 2003, \apjs, 147, 197

\bibitem[{{Heckman} {et~al.}(1990){Heckman}, {Armus}, \&
  {Miley}}]{Heckmanetal1990}
{Heckman}, T.~M., {Armus}, L., \& {Miley}, G.~K. 1990, \apjs, 74, 833

\bibitem[{{Hopkins} {et~al.}(2011){Hopkins}, {Quataert}, \&
  {Murray}}]{Hopkinsetal2011}
{Hopkins}, P.~F., {Quataert}, E., \& {Murray}, N. 2011, \mnras, 417, 950

\bibitem[{{Hopkins} {et~al.}(2012){Hopkins}, {Quataert}, \&
  {Murray}}]{Hopkinsetal2012}
---. 2012, \mnras, 421, 3522

\bibitem[{{Jacquet} \& {Krumholz}(2011)}]{JacquetKrumholz2011}
{Jacquet}, E., \& {Krumholz}, M.~R. 2011, \apj, 730, 116

\bibitem[{{Jiang} {et~al.}(2013){Jiang}, {Davis}, \& {Stone}}]{Jiangetal2013}
{Jiang}, Y.-F., {Davis}, S.~W., \& {Stone}, J.~M. 2013, \apj, 763, 102

\bibitem[{{Jiang} {et~al.}(2012){Jiang}, {Stone}, \& {Davis}}]{Jiangetal2012}
{Jiang}, Y.-F., {Stone}, J.~M., \& {Davis}, S.~W. 2012, submitted to ApJS

\bibitem[{{Kennicutt}(1998)}]{Kennicutt1998}
{Kennicutt}, Jr., R.~C. 1998, \apj, 498, 541

\bibitem[{{Krolik}(1977)}]{Krolik1977}
{Krolik}, J.~H. 1977, Physics of Fluids, 20, 364

\bibitem[{{Krumholz} {et~al.}(2007){Krumholz}, {Klein}, {McKee}, \&
  {Bolstad}}]{Krumholzetal2007}
{Krumholz}, M.~R., {Klein}, R.~I., {McKee}, C.~F., \& {Bolstad}, J. 2007, \apj,
  667, 626

\bibitem[{{Krumholz} \& {Matzner}(2009)}]{KrumholzMatzner2009}
{Krumholz}, M.~R., \& {Matzner}, C.~D. 2009, \apj, 703, 1352

\bibitem[{{Krumholz} \& {Thompson}(2012)}]{KrumholzThompson2012}
{Krumholz}, M.~R., \& {Thompson}, T.~A. 2012, \apj, 760, 155

\bibitem[{{Krumholz} \& {Thompson}(2013)}]{KrumholzThompson2013}
---. 2013, \mnras, 434, 2329

\bibitem[{{Kunasz} \& {Auer}(1988)}]{KunaszAuer1988}
{Kunasz}, P., \& {Auer}, L.~H. 1988, \jqsrt, 39, 67

\bibitem[{{Larsen} \& {Wollaber}(2008)}]{LarsenWollaber2008}
{Larsen}, E.~W., \& {Wollaber}, A.~B. 2008, Nuclear Science and Engineering,
  160, 267

\bibitem[{{Levermore} \& {Pomraning}(1981)}]{LevermorePomraning1981}
{Levermore}, C.~D., \& {Pomraning}, G.~C. 1981, \apj, 248, 321

\bibitem[{{Lowrie} {et~al.}(1999){Lowrie}, {Morel}, \&
  {Hittinger}}]{Lowrieetal1999}
{Lowrie}, R.~B., {Morel}, J.~E., \& {Hittinger}, J.~A. 1999, \apj, 521, 432

\bibitem[{{Mathews} \& {Blumenthal}(1977)}]{MathewsBlumenthal1977}
{Mathews}, W.~G., \& {Blumenthal}, G.~R. 1977, \apj, 214, 10

\bibitem[{{Mihalas}(1978)}]{Mihalas1978}
{Mihalas}, D. 1978, {Stellar atmospheres /2nd edition/}

\bibitem[{{Mihalas} \& {Klein}(1982)}]{MihalasKlein1982}
{Mihalas}, D., \& {Klein}, R.~I. 1982, Journal of Computational Physics, 46, 97

\bibitem[{{Mihalas} \& {Mihalas}(1984)}]{MihalasMihalas1984}
{Mihalas}, D., \& {Mihalas}, B.~W. 1984, {Foundations of radiation
  hydrodynamics}, ed. {Mihalas, D.~\& Mihalas, B.~W.}

\bibitem[{{Murray} {et~al.}(2005){Murray}, {Quataert}, \&
  {Thompson}}]{Murrayetal2005}
{Murray}, N., {Quataert}, E., \& {Thompson}, T.~A. 2005, \apj, 618, 569

\bibitem[{{Pettini} {et~al.}(2001){Pettini}, {Shapley}, {Steidel}, {Cuby},
  {Dickinson}, {Moorwood}, {Adelberger}, \& {Giavalisco}}]{Pettinietal2001}
{Pettini}, M., {Shapley}, A.~E., {Steidel}, C.~C., {Cuby}, J.-G., {Dickinson},
  M., {Moorwood}, A.~F.~M., {Adelberger}, K.~L., \& {Giavalisco}, M. 2001,
  \apj, 554, 981

\bibitem[{{Rupke} {et~al.}(2005){Rupke}, {Veilleux}, \&
  {Sanders}}]{Rupkeetal2005}
{Rupke}, D.~S., {Veilleux}, S., \& {Sanders}, D.~B. 2005, \apjs, 160, 115

\bibitem[{{Schwartz} \& {Martin}(2004)}]{SchwarzMartin2004}
{Schwartz}, C.~M., \& {Martin}, C.~L. 2004, \apj, 610, 201

\bibitem[{{Scoville} {et~al.}(2001){Scoville}, {Polletta}, {Ewald}, {Stolovy},
  {Thompson}, \& {Rieke}}]{Scoville2001}
{Scoville}, N.~Z., {Polletta}, M., {Ewald}, S., {Stolovy}, S.~R., {Thompson},
  R., \& {Rieke}, M. 2001, \aj, 122, 3017

\bibitem[{{Sekora} \& {Stone}(2010)}]{SekoraStone2010}
{Sekora}, M.~D., \& {Stone}, J.~M. 2010, Journal of Computational Physics, 229,
  6819

\bibitem[{{Semenov} {et~al.}(2003){Semenov}, {Henning}, {Helling}, {Ilgner}, \&
  {Sedlmayr}}]{Semenovetal2003}
{Semenov}, D., {Henning}, T., {Helling}, C., {Ilgner}, M., \& {Sedlmayr}, E.
  2003, \aap, 410, 611

\bibitem[{{Shaviv}(2001)}]{Shaviv2001}
{Shaviv}, N.~J. 2001, \apj, 549, 1093

\bibitem[{{Shestakov} \& {Offner}(2008)}]{ShestakovOffner2008}
{Shestakov}, A.~I., \& {Offner}, S.~S.~R. 2008, Journal of Computational
  Physics, 227, 2154

\bibitem[{{Socrates} \& {Sironi}(2013)}]{SocratesSironi2013}
{Socrates}, A., \& {Sironi}, L. 2013, \apjl, 772, L21

\bibitem[{{Stone} {et~al.}(2008){Stone}, {Gardiner}, {Teuben}, {Hawley}, \&
  {Simon}}]{Stoneetal2008}
{Stone}, J.~M., {Gardiner}, T.~A., {Teuben}, P., {Hawley}, J.~F., \& {Simon},
  J.~B. 2008, \apjs, 178, 137

\bibitem[{{Thompson} {et~al.}(2005){Thompson}, {Quataert}, \&
  {Murray}}]{Thompsonetal2005}
{Thompson}, T.~A., {Quataert}, E., \& {Murray}, N. 2005, \apj, 630, 167

\bibitem[{{Wise} {et~al.}(2012){Wise}, {Abel}, {Turk}, {Norman}, \&
  {Smith}}]{Wiseetal2012}
{Wise}, J.~H., {Abel}, T., {Turk}, M.~J., {Norman}, M.~L., \& {Smith}, B.~D.
  2012, \mnras, 427, 311

\end{thebibliography}
\end{document}